\documentclass[12pt]{iopart}

\usepackage[perpage,marginal]{footmisc}
\makeatletter
\renewcommand\footnoterule{%
  \kern-3\p@
  \hrule\@width2.5cm
  \kern2.6\p@}
\makeatother

\usepackage{color, xcolor}
\usepackage[hypertexnames=false, unicode, colorlinks, allcolors=blue!70!black, linktocpage, pdfusetitle]{hyperref}
\usepackage[all]{hypcap}
\usepackage{amssymb,amsbsy,amsfonts}
\usepackage{indentfirst}
\usepackage[pdftex]{graphicx}
\usepackage[numbers,square,comma,sort&compress]{natbib}
\usepackage{url}

\usepackage[T1]{fontenc}
\usepackage[utf8]{inputenc}
\usepackage{lmodern}

\usepackage{dcolumn}
\usepackage{bm}
\usepackage{hyperref}
\usepackage{etoolbox}

\usepackage{orcidlink}

\usepackage{caption}
\usepackage[most]{tcolorbox}
\usepackage{subcaption}
\usepackage{float}
\usepackage{enumitem}
\newlist{steps}{enumerate}{1}
\setlist[steps, 1]{label = Step \arabic*:}

\begin{document}

\title[Infinitely degenerate slowly rotating solutions in $f(R)$ gravity]{Infinitely degenerate slowly rotating solutions in $f(R)$ gravity}

\author{Alan Sunny$^{1}$, Semin Xavier\orcidlink{0000-0003-2025-5345}$^2$ and S. Shankaranarayanan\orcidlink{0000-0002-7666-4116}$^3$}
\address{Department of Physics, Indian Institute of Technology Bombay, Mumbai 400076, India}
 \ead{$^1$alansunny21.phy@gmail.com,\\~~~~~~ $^2$seminxavier@iitb.ac.in,\\~~~~~~~$^3$shanki@iitb.ac.in } 

\begin{abstract}
This work tests the no-hair conjecture in $f(R)$ gravity models. No-hair conjecture asserts that all black holes in General Relativity coupled to any matter must be Kerr–Newman type. However, the conjecture fails in some cases with non-linear matter sources. Here, we address this by explicitly constructing multiple slow-rotating black hole solutions, up to second order in rotational parameter, for a class of $f(R)$ models ($f(R) =(\alpha_{0} + \alpha_{1}\,R)^{p}, p > 1$). Such an $f(R)$ includes all higher-powers of $R$. We analytically show that multiple vacuum solutions satisfy the field equations up to the second order in the rotational parameter. In other words, we show that the multiple vacuum solutions depend on arbitrary constants, which depend on the coupling parameters of the model. Hence, our results indicate that the no-hair theorem for modified gravity theories merits extending to include the coupling constants. 
The uniqueness of our result stems from the fact that these are obtained directly from metric formalism without conformal transformation. We discuss the kinematical properties of these black hole solutions and compare them with slow-rotating Kerr. Specifically, we show that the circular orbits for the black holes in $f(R)$ are smaller than that of Kerr. This implies that the inner-most stable circular orbit for black holes in $f(R)$ is smaller than Kerr's; hence, the shadow radius might also be smaller. Finally, we discuss the implications of our results for future observations.
\end{abstract}
\maketitle
\normalsize
\clearpage

\section{Introduction}
\label{sec:Intro}

The direct observation of gravitational waves (GWs) from a binary black hole (BH) merger in 2015 opened a new frontier for investigating the event horizon of the residual BH, where the curvature scale can be as large as $10^{-2}\, {\rm km}^{-1}$~\cite{2016-Abbott.etal-PRL,2021-LIGO-TestGR-PRD,2021-GWTC2-TestGR-Arxiv}. However, the validity of General Relativity (GR) has yet to be rigorously investigated in these strong-gravity regimes~\cite{2008-Psaltis-LRR,2018-Berti.etal-GRGa,2018-Berti.etal-GRGb,2019-Barack.etal-CQG,2022-Shanki.Joseph-GRG}. Observational cosmology's biggest surprise on the largest scale is the accelerated expansion of the current Universe~\cite{1998-Riess.Others-Astron.J.}. This can be explained by the presence of a mysterious energy source known as dark energy or by modifications to GR on the cosmological scales~\cite{2003-Peebles.Ratra-RMP,2003-Padmanabhan-Phys.Rept.,2010-DeFelice.Tsujikawa-LivingRev.Rel.,2010-Sotiriou.Faraoni-Rev.Mod.Phys.,2012-Clifton.etal-Phys.Rept.,2016-Joyce.etal-ARN,2017-Nojiri.etal-Phys.Rept.,2021-Riess.others}.

LIGO-VIRGO-KAGRA observations confirmed GR's predictions and enriched our understanding of the Universe by providing the first direct evidence of massive stellar-mass BHs and BHs colliding to form a single, larger BH. The existence of BH singularities, as described by GR, suggests that the theory may not accurately describe the universe at extremely small scales. This suggests that while GR is an extremely successful theory over large scales, it may not accurately describe the smallest scales or strong gravity. One possibility is to modify GR in the strong gravity limit by introducing higher derivative Ricci scalar, Ricci tensor, and Riemann tensor terms in action~\cite{2007-Woodard-Lect.NotesPhys.,2010-DeFelice.Tsujikawa-LivingRev.Rel.,2010-Sotiriou.Faraoni-Rev.Mod.Phys.,2012-Clifton.etal-Phys.Rept.,2016-Joyce.etal-ARN,2017-Nojiri.etal-Phys.Rept.}. Unlike 4-dimensional GR, whose field equations contain only up to second-order derivatives, the modified theories with higher derivative Ricci/Riemann tensor gravity models include higher derivatives~\cite{1983-Barth-Christensen-PRD,1990-Simon-PRD,1991-Mueller-Hoissen-AnP,2001-Hawking.Hertog-PRD}. Therefore, one expects significant differences between GR and modified theories in the strong-gravity regime~\cite{1991-Simon-PRD,1992-Simon-PRD}. 

Currently, two approaches are employed to distinguish GR from modified gravity theories in Electromagnetic and GW observations: First, identify modified gravity theories with the same BH solutions as GR. Then, obtain the difference in the GW signals from these two theories~\cite{2017-Bhattacharyya.Shanki-PRD,2018-Bhattacharyya.Shanki-EPJC,2019-Bhattacharyya.Shanki-PRD,2019-Shanki-IJMPD,2022-Chowdhury.etal-Arxiv}. Second, obtain new BH solutions for modified gravity theories~\cite{2009-Yunes.Pretorius-PRD,2012-Yagi.etal-PRD,2015-Maselli.etal-PRD,2019-Cano.etal-JHEP}. Then, use template matching to match GW signals and identify the deviations, if any~\cite{2019-Isi.etal-PRL,2018-Baibhav.etal-PRD}. In this work, we follow the second approach. There are two key reasons for this.

First, according to Birkhoff's theorem, the Schwarzschild solution is the unique spherically symmetric vacuum solution of the Einstein field equations. Recently, two current authors obtained an infinite number of exact static spherically symmetric vacuum solutions for a class of $f(R)$ gravity ($f(R) =(\alpha_{0} + \alpha_{1}\,R)^{p}, p > 1$)~\cite{2020-Xavier.etal-CQG}. It was explicitly shown that the \emph{Birkhoff theorem} is not valid for all modified gravity theories. 
In GR, the zero-spin ($J \to 0$) limit of the Kerr BH uniquely leads to the Schwarzschild solution. Thus, if a large number of spherically symmetric vacuum solutions exist in $f(R)$, the rotating solution also \emph{may not} be unique. 

Second, there is no uniqueness theorem in GR for space-time describing a rotating star. Hence, 
it is reasonable to assume that the stationary axisymmetric BH solutions are a large family. Remarkably, all stationary, asymptotically flat, vacuum solutions of the Einstein field equations that contain a regular horizon with no singularities outside the horizon are given by a two-parameter family~\cite{1985-Chandrasekhar-Book,1973-Misner.etal-Book}. In the case of GR, we know one such family --- the Kerr --- is unique~\cite{2012-Chrusciel.etal-LRR}. Due to \emph{no hair theorem}, the only memory of the nature, structure, and composition of any object that collapses to form a stationary BH is embodied in the mass ($M$) and angular momentum ($a$), with any residual hair rapidly radiated away during the collapse process. This raises a few questions: Is the no-hair theorem a feature of gravity or GR\footnote{It has been shown that the \emph{standard} no-hair theorem fails in some cases, like Einstein–Yang–Mills theory and nonlinear electrodynamics, among others~\cite{2015-Herdeiro.Radu-IJMPD,2016-Barrientos.etal-EPJC}.}? In other words, do modified gravity theories support other rotating BH solutions besides Kerr? If yes, which class of modified gravity theories supports and how are they different from the Kerr solution in GR? Do we need to enhance the no-hair theorem for BHs in modified gravity theories? 

This work addresses some of these questions by explicitly constructing multiple slow rotating black hole (SRBH) solutions for a class of $f(R)$ gravity models ($f(R) =(\alpha_{0} + \alpha_{1}\,R)^{p}, p > 1$). Unfortunately, modified theories typically lead to more complex field equations, and obtaining exact rotating BH solutions analytically is challenging. Unlike GR, whose field equations contain only up to second-order derivatives, the equations of $f(R)$ gravity models include higher derivatives. Hence, the higher-order derivative corrections, with a significant number of degrees of freedom,  cannot be treated as a perturbation to GR, especially in the strong-gravity regime~\cite{2022-Shanki.Joseph-GRG}. However small they may appear in the Newtonian limit, higher-derivative terms make the new theory drastically different from GR~\cite{1983-Barth-Christensen-PRD,1990-Simon-PRD,1991-Mueller-Hoissen-AnP,2001-Hawking.Hertog-PRD,1991-Simon-PRD,1992-Simon-PRD}.

In GR, no-hair theorems concern mostly black hole solutions with flat asymptotics. It is known that hairy black holes with non-flat asymptotics often exist. For instance, non-linear field sources, such as Yang-Mills fields~\cite{1989-Volkov-JETP,1999-Volkov-PhyRept,1990-Bizon-PRL,1993-Greene-PRD} and non-canonical scalars such as Skyrmions \cite{1986-Hugh_Ian-PLB,1991-Markus_Serge_NorbertStraumann-PLB} lead to hairy black-holes. Recently, it has been shown that scalar-tensor Horndeski theories with higher derivatives can lead to spherically symmetric hairy BHs~\cite{2014-Herdeiro-PRL,2014-Babichev-JHEP,2014-Sotiriou_Zhou-PRD,2014-Charmousis_Papantonopoulos-JHEP}. Also, it has been shown that massive, minimally-coupled complex scalar fields can lead to hairy Kerr solutions~\cite{2014-Herdeiro-PRL}. {For completeness, we have provided more details about the no hair violation in other theories of gravity in \ref{sec:refs}.}

{To our knowledge, exact (analytical) rotating black hole solutions have not been obtained in the literature for (higher-derivative) modified gravity theories in 4-D.} For instance, in the case of dynamical Chern-Simons gravity, the presence of axial scalar field through the Pontryagin density leads to a non-Kerr rotating black hole~\cite{2009-Yunes_Pretorius-PRD}. In the case of Chern-Simons gravity, the deviation from the Kerr depends on the Chern-Simons coupling constant. Note that this solution is asymptotically flat and is obtained up to the second order in rotational parameter $\chi (= a/M) = J/M^2$, $J$ is the angular momentum, and $M$ is the mass). 

In this work, to keep the calculations tractable, we analytically obtain {asymptotically non-flat}, SRBH solutions up to second-order in the rotational parameter ($\chi$)  for $f(R)$ gravity. We explicitly show that a class of $f(R)$ model leads to multiple SRBH solutions. The results are then compared with Kerr and SR Kerr solutions (in GR) by evaluating different physical parameters. We also discuss the implications of our result in the conformally equivalent frame.

The rest of this work is organized as follows: {In Sec.~\ref{sec:KerrSchild}, we discuss the generic $f(R)$ gravity and the existence of multiple rotating black hole solutions for $f(R)$ gravity using the Kerr-Schild decomposition of the 4-D metric.}
In Sec.~\ref{sec:Solutions}, we discuss the $f(R)$ model and the procedure we followed to obtain the slowly-rotating black hole solutions up to second-order in $\chi$. In Sec.~\ref{sec:Properties}, we discuss the properties of these solutions and discuss the difference w.r.t Kerr solution in GR. Finally, in Sec.~\ref{sec:conc}, we summarize our results and discuss possible future directions. Appendices contain details of the calculations. In this work, we use $(-, +, +, +)$ metric signature and set $G = c = 1$.  Greek alphabets denote the 4-D space-time coordinates. A prime denote derivative w.r.t $\rho$. 


\section{$f(R)$ and the case for multiple rotating black hole solutions}
\label{sec:KerrSchild}

In this section, we discuss the generic $f(R)$ gravity. Later, using the Kerr-Schild decomposition of the 4-D metric, we show that multiple rotating black hole solutions exist for $f(R)$ gravity. 

\subsection{{\it f(R)} gravity}

The modified $f(R)$ gravity action in strong-gravity is:
\begin{equation}
\begin{aligned}
 S &= \frac{1}{16 \pi}\int d^4x \, \sqrt{-g} \, f(R) \\
  &= \frac{1}{16 \pi}\int d^4x \, \sqrt{-g} \, \left[c_{0} + c_{1}\,R + c_{2}\,R^2 + \cdots \right]
\label{eq:fRact-Jordan}
\end{aligned}
\end{equation}
To avoid instabilities and ghosts, the coefficients $c_0, c_1, c_2, \cdots$ must satisfy $\partial f / \partial R > 0, \partial^2 f /\partial R^2 > 0$~\cite{2022-Shanki.Joseph-GRG}. Using metric formalism, the modified Einstein's equation is:
%
\begin{align}\label{ModifiedFEq}
&{\cal G}_{\mu\nu} \equiv  \partial_R[f(R)] R_{\mu\nu}-\frac{1}{2}f(R)\,g_{\mu\nu} - \left[\nabla_{\mu}\nabla_{\nu} - g_{\mu\nu}\,\Box \right] \partial_R[f(R)] = 0 
\end{align}
%
where $\Box = \nabla^{\mu}\,\nabla_{\mu}$. Taking 
covariant differentiation of the above equation leads to the generalized Bianchi identity~\cite{2016-Tian-GRG,2019-Shanki-IJMPD}:
\begin{align}
    \partial_R^2[f(R)] \, R_{\mu\nu} \nabla^{\mu}R = 0 \, . \label{GeneralizedBianchi}
\end{align}
The generalized Bianchi identity leads to four constraints on the Ricci tensor. For GR ($f(R) = R$), $\partial_R^2[f(R)] = 0$ and the above condition is trivially satisfied.

Before we proceed with the evaluation of non-trivial axisymmetric BH solutions, we want to highlight why it is natural to expect non-Kerr solutions in $f(R)$ gravity. To see this, let us rewrite Eq. \eqref{ModifiedFEq} in the following form:
\begin{align}\label{ModifiedFEq2}
\partial_R[f(R)] R_{\mu\nu} = \frac{1}{2}f(R)\,g_{\mu\nu} + \left[\nabla_{\mu}\nabla_{\nu} - g_{\mu\nu}\,\Box \right] \partial_R[f(R)] 
\end{align}
By definition, the Kerr solution satisfies the condition $R_{\mu\nu} = 0$ and (hence, $R = 0$). In other words, any $f(R)$ gravity model that satisfies $R_{\mu \nu} =0$ will have Kerr metric as a black hole solution. 
However, no-hair theorem in GR 
states that black-holes are \emph{uniquely} described by the Kerr metric. In order for Kerr to be the  unique solution of the above modified Einstein's equation, $f(R)$ must satisfy the following condition:
\begin{align}\label{ModifiedFEq3}
 \frac{1}{2}f(R)\,g_{\mu\nu} + \left[\nabla_{\mu}\nabla_{\nu} - g_{\mu\nu}\,\Box \right] \partial_R[f(R)] = 0 
\end{align}
Any arbitrary function of $f(R)$ will not satisfy the above condition for all coordinates ($x^{\mu}$) since the coupling constants are arbitrary. In the following section, using Kerr-Schild decomposition we 
show this is indeed the case. 
\subsection{Rotating black hole solution in Kerr-Schild form}
\label{subsec:KerrSchild}
{
In GR, the rotating --- Kerr and Kerr-Newman --- solutions allow a Kerr-Schild decomposition of the metric~\cite{2009-Kerr.Schild-GRG,2018-Deruelle-jeanUzan-book}:}
\begin{equation}
\label{eq:KerrSchilddecom}
    g_{\mu \nu}=\bar{g}_{\mu \nu}+l_{\mu \nu} \quad \text{with}\quad l_{\mu \nu}=h(x^{\rho})l_{\mu}l_{\nu} 
    \end{equation}
{where $\bar{g}_{\mu \nu}$ is the seed (or known 'background') space-time, $h(x^{\rho})$ is an a priori arbitrary
function of the space-time coordinates and the vector field $l^{\mu} = \bar{g}^{\mu \nu} l_{\nu} $ is null $\bar{g}_{\mu \nu} l^{\mu}  l^{\nu}=0$. The vector field $l^{\mu}$ satisfies the geodesic equation
$l^{\nu}\bar{D}_{\nu}l^{\mu}=0$, where $\bar{D}$ is the covariant derivative w.r.t $\bar{g}_{\mu \nu}$. For the above decomposition \eqref{eq:KerrSchilddecom}, the Ricci tensor and Ricci scalar can be rewritten as:}
\begin{align}
     R^{\mu}_{\nu}&=\bar{R}^{\mu}_{\nu}-l^{\mu\rho} \bar{R}_{\rho \nu}+\bar{D}_{\rho}(\bar{g}^{\mu \lambda}  \Delta^{\rho}_{\nu \lambda} ) \quad  \text{where} \quad \Delta^{\mu}_{\nu \rho}=\frac{1}{2}\left( \bar{D}_{\nu} l^{\mu}_{\rho} +\bar{D}_{\rho } l^{\mu}_{\nu}-\bar{D}^{\mu} l_{\nu \rho} \right)\label{kerr-schildRicciT}
     \\
       R &= \bar{R}- l^{\mu \nu}\bar{R}_{\mu \nu}+\bar{D}_{\mu}{\cal V}^{\mu} \quad ~~~~~~\quad     \text{where} \quad {\cal V}^{\mu}=l^{\mu}\bar{D}_{\nu}(h \, l^{\nu})\label{kerr-schildRicciS}
\end{align}
{
More specifically, Kerr-Schild showed that the rotating black hole metric in GR can be constructed by using the Minkowski metric, in spheroidal coordinates, as seed metric:}
\begin{equation}\label{eqn:sedmetrc}
    ds^2=-dT^2+\frac{(r^2 +a^2 \cos^2 \theta)}{(r^2 + a^2)}dr^2+(r^2 +a^2 \cos^2 \theta)d\theta^2+(r^2 + a^2)\sin^2\theta d\phi^2
\end{equation}
{
where $a$ is a constant. We now assume that the rotating black hole solutions in $f(R)$ allow a Kerr-Schild decomposition \eqref{eq:KerrSchilddecom}. Substituting Eqs. \eqref{kerr-schildRicciT}, \eqref{kerr-schildRicciS} in Eq. \eqref{GeneralizedBianchi}, we have:}
\begin{equation}\label{fR-KerrschildCon-fildeq}
  \partial_R^2[f(R)] \,\left[\bar{R}_{\mu\nu}-l^{\rho}_{\mu} \bar{R}_{\rho \nu}+\bar{D}_{\rho}(\bar{g}^{ \lambda}_{\mu} \Delta^{\rho}_{\nu \lambda}) \right] \bar{D}^{\mu}\left[ \bar{R}- l^{\rho \sigma}\bar{R}_{\rho \sigma}+\bar{D}_{\rho}{\cal V}^{\rho} \right] = 0 .  
\end{equation}
{For the seed metric \eqref{eqn:sedmetrc}, $\bar{R}_{\mu\nu} =0$, $\bar{R} = 0$ and the null geodesic vector is $ l^{\mu}=(1,-1,0,\frac{a}{a^2 +r^2})$. Substituting these in Eq.~\eqref{fR-KerrschildCon-fildeq}, we have:}
\begin{equation}
 \partial_R^2[f(R)] \, \bar{D}_{\rho}\left[\bar{g}^{ \lambda}_{\mu} \Delta^{\rho}_{\nu \lambda}\right] \, \bar{D}^{\mu}\left[\bar{D}_{\rho}{\cal V}^{\rho} \right] = 0.  
\end{equation}
{The above equation is trivially satisfied for GR because $\partial_R^2[f(R)] = 0$. In the case of $f(R)$, $\partial_R^2[f(R)]$ is non-zero. Hence, the above equation leads to two conditions: }
\[
\bar{D}_{\rho}(\bar{g}^{ \lambda}_{\mu} \Delta^{\rho}_{\nu \lambda})=0~~\mbox{or}~~\bar{D}^{\mu}\left( \bar{D}_{\rho}{\cal V}^{\rho} \right)=0 \, . 
\]
We thus see at least two rotating solutions exist in $f(R)$. One solution that satisfies the first condition and the second solution that satisfies the second one. This is because both conditions satisfy distinct second-order differential equations.

{Using the symmetry of the rotating solutions, we assume $h(x^\rho)$ is only a function of $r$ and $\theta$. The $r$-component of the second condition $ { \bar{D}^{\mu}\left( \bar{D}_{\rho}{\cal V}^{\rho} \right) }=0$, leads to:}
\begin{equation}
h(r,\theta)=\frac{1}{(r^2 +a^2 \cos^2\theta)}\left( 
{\cal H}_0 (\theta) + {\cal H}_1(\theta)r + {\cal H}_2(\theta) r^2 +  {\cal H}_4 r^4 \right) 
\end{equation}
{where ${\cal H}_0(\theta)$ and ${\cal H}_1(\theta)$ are arbitrary functions, ${\cal H}_2(\theta) = a^2 \cos^2(\theta)/2$ and ${\cal H}_4(\theta) = 1/12$.}

This is the key result of this section, regarding which we want to discuss the following points: First, we have explicitly shown that any $f(R)$ gravity can have at least two rotating black hole solutions. Second, the analysis rests on the fact that the rotating black holes allow a Kerr-Schild form. At the same time, this may be considered as a restriction, as in the case of GR, the above solutions can be transformed to Boyer-Lindquist form. Third, in the above analysis, we have assumed the seed metric to have a vanishing Ricci tensor. In the rest of the work, we relax this assumption and show that the above result also holds for Ricci flat space-times. This leads to the following question: What is the implication of the multiple rotating black hole solutions for $f(R)$ theories of gravity? Let us focus on ${\cal H}_i(\theta) $'s to answer this. In GR, \emph{only }${\cal H}_1(\theta)$ is non-vanishing and all other coefficients vanish. However, in the $f(R)$, these are non-vanishing and contain information about the form of $f(R)$. More specifically, the coefficients may depend on the coupling constants. In other words, the above analysis suggests that the no-hair theorem for $f(R)$ should include the coupling constants. In the 
rest of this we explicit show this by considering a specific form $f(R) =(\alpha_{0} + \alpha_{1}\,R)^{p}, p > 1$. Although the analysis can be done for any $p > 1$, to keep things transparent, we consider $p = 2$.

\section{Multiple slowly-rotating black-hole solutions} 
\label{sec:Solutions}

In this section, we obtain black solutions in Ricci-flat space-times. However, it is not possible to obtain exact rotating solution for arbitrary $f(R)$ models. To keep the calculations tractable, we assume $f(R)$ in the following binomial form: 
\begin{align}
\label{f(R)2ndorder}
    f(R) =(\alpha_{0} + \alpha_{1}\,R)^{p}
\end{align}
where $\alpha_{0}, \alpha_{1} > 0$ and $p > 1$. Here are some points we want to discuss about the action: First, the coefficient $\alpha_{0}$ is dimensionless, and $\alpha_{1}$ has a dimension of $[L]^2$. Second, in the limit, $\kappa^2 \equiv \alpha_1/\alpha_0 \ll 1$, the above action reduces to GR with a cosmological constant. However, both these coefficients have to be non-zero. Third, for the most part, we obtain solutions for $p = 2$; however, the results can be extended to any $p$. Substituting the above form of $f(R)$ in Eq.~\eqref{ModifiedFEq}, we have:
\begin{equation}
\begin{aligned}
G_{\mu\nu} &= - \kappa^2 R \left(R_{\mu\nu} - \frac{1}{4} g_{\mu\nu} R \right)
+ \kappa^2 \left[\nabla_{\beta}\nabla_{\gamma} - g_{\beta \gamma}\,\Box \right] R
+ \frac{1}{4 \kappa^2} g_{\mu\nu} 
\label{SecondOrderModified-General}
\end{aligned}
\end{equation}
As it is evident the equations of motion are higher-order and hence, obtaining exact rotating black hole solutions is non-trivial. We aim to construct a non-trivial slowly-rotating black hole (SRBH) solutions up to second-order in $\chi$ 
for the above $f(R)$ gravity.

In this work, we extend Hartle's approach~\cite{1967-Hartle-ApJ,1968-Hartle.Thorne-ApJ} in which spin corrections to the static spherically symmetric solutions are introduced perturbatively. This approach has recently been applied to obtain SRBHs in Chern-Simons and Dilaton-Gauss-Bonnet gravity~\cite{2012-Yagi.etal-PRD,2015-Maselli.etal-PRD}.

\subsection{Multiple SRBH solutions in {\it f(R)} gravity}

As shown in \ref{App:A}, the first order $(\chi)$ SR solution for any $p$ is highly degenerate (see Eq. \eqref{eq:Linearorder-Bianchi}). Hence, in the rest of this work, we consider the following line-element that corresponds to a slowly-rotating space-time up to quadratic in the spin parameter $(\chi)$: 
%

{\small
\begin{align} 
\label{Slow_Kerr}
 ds^2 & = - e^{\mu(\rho)} \left[U(\rho) + \chi^2 \, V(\rho) \cos^2(\theta)\right] d\tau^2 
+ \frac{1}{U(\rho)}\left[1 - \chi^2\left(\frac{V(\rho)}{U(\rho)} + \frac{\sin^2(\theta)}{\rho^2}\right)\right]\,d\rho^2 
\\ 
&- 2\,\chi\, V(\rho)\rho^2 \sin^2(\theta)\,d\tau\,d\phi + \left[\rho^2 + \chi^2\, \cos^2(\theta)\right]d\theta^2 + \rho^2 \sin^2(\theta)\,\left[1 + \frac{\chi^2}{\rho^2}+\chi^2\,V(\rho) \sin^2(\theta)\right]\,d\phi^2 \nonumber
\end{align}
}
%

where $U(\rho), V(\rho)$ and $\mu(\rho)$ are unknown functions of the new radial coordinate $\rho$. Demanding $g_{\phi\phi}$ is positive definite implies that $V(\rho) > 0$. Similarly we demand that $\exp[\mu(\rho)]$ is non-zero for all $\rho$. $\tau, \rho$ are dimensionless time ($t$) and radial coordinates ($r$): $\tau = {t}/{M}, \rho = {r}/{M}$. A word of note regarding $M$: We will identify one of the unknown coefficients with the black hole mass; until then, $M$ has no physical meaning. 

To obtain solutions, we group terms corresponding to different orders in $\chi$~\cite{1967-Hartle-ApJ,1968-Hartle.Thorne-ApJ}, i. e.,
\small{
\begin{equation}
{\cal G} \simeq f^{(0)}(\rho) + 
f^{(I)}(\rho, \theta)\, \chi + 
f^{(II)}(\rho, \theta) \, \chi^2 + 
{\cal O}(\chi^3) + \cdots \, ,
\end{equation}
}
where $f^{(0)}(\rho)$ corresponds to spherically symmetric metric. More specifically, substituting the line element \eqref{Slow_Kerr} in Eq.~\eqref{SecondOrderModified-General},  eliminating terms greater than $\chi^2$, leads to: 
\small{
\begin{subequations}
\label{eq:Secord-ModEqs}
\begin{align}
    {\cal G}^{\tau}_{\tau} \equiv T_{1}[U(\rho),V(\rho),\mu(\rho)] &\simeq T_{1}^{(0)} (\rho) + T^{(II)}_{1} (\rho, \theta) \chi^2  = 0\label{eq:Secord-ModEqsa}\\
    {\cal G}^{\rho}_{\rho} \equiv T_{2}[U(\rho),V(\rho),\mu(\rho)] &\simeq T_{2}^{(0)} (\rho) + T^{(II)}_{2} (\rho, \theta) \chi^2 = 0\label{eq:Secord-ModEqsb}\\
    {\cal G}^{\theta}_{\theta} \equiv T_{3}[U(\rho),V(\rho),\mu(\rho)] &\simeq T_{3}^{(0)} (\rho) + T^{(II)}_{3} (\rho, \theta) \chi^2  = 0\label{eq:Secord-ModEqsc}\\
     {\cal G}^{\rho}_{\theta} \equiv T_{4}[U(\rho),V(\rho),\mu(\rho)] &\simeq T_{4}^{(II)}(\rho, \theta) \, \chi^2 = 0\label{eq:Secord-ModEqsd}\\
    {\cal G}^{\tau}_{\phi} \equiv T_{5
    }[U(\rho),V(\rho),\mu(\rho)] &\simeq T_{5}^{(I)}(\rho, \theta) \, \chi = 0 \label{eq:Secord-ModEqse}\, ,
\end{align}
\end{subequations}
}
where, $T_i$ ($i = 1 \cdots 5$) refers to the modified Einstein tensor component containing terms up to second order in $\chi$. $T_i^{(0)}(\rho)$ corresponds to
modified Einstein tensor components of the spherically symmetric part while $T_i^{(II)}(\rho, \theta)$ corresponds to
modified Einstein tensor components of the second order $\chi$. This is the key equation regarding which we want to mention the following points:
\begin{enumerate}
\item  $T_{1}$ and $T_{3}$ contain fourth-order derivatives of $U(\rho), \mu(\rho)$ and $V(\rho)$, while $T_{2}$ and $T_4$ contain third-order derivatives of $U(\rho), \mu(\rho)$ and $V(\rho)$. $T_{5}$ contains third-order derivatives of $U(\rho)$ and $\mu(\rho)$, and second-order derivative of $V(\rho)$. Explicit expressions are not illuminating, so we do not report them here. \href{https://www.dropbox.com/sh/5cwqegs6063lkz3/AAC4b2dxUL84zgoQKeMZxgoMa?dl=0}{Details are in the online MAPLE files.}
\item $T_1, T_2$ and $T_3$ do not contain terms in first-order in $\chi$, i.e.
$T_{i}^{(0)} (\rho) + T^{(II)}_{i} (\rho) \chi^2 +{\cal O}(\chi^3)$, where $i=1,2$ and $3$. Additionally, these three components' $\chi$ independent terms depend only on $U(\rho)$ and $\mu(\rho)$.
\item $T_4$ only contains second-order in $\chi$, i. e. $T_{4}^{(II)}(\rho) \,  \chi^2 + {\cal O}(\chi^3)$, while $T_5$ only contains first-order in $\chi$, i. e, 
$T_{5}^{(II)}(\rho) \, \chi + {\cal O}(\chi^3)$. 
\item It is possible to express $ {\cal G}^{\phi}_{\tau}$ and $ {\cal G}^{\theta}_{\rho}$ in terms of $T_{1}, T_2, T_3, T_4$ and $T_5$. These are consistent with the results in \ref{App:A}.
\item Since the equations of motion contain fourth-order derivatives of $U(\rho)$, $V(\rho)$, and $\mu(\rho)$, an exact solution will have four independent constants. 
\end{enumerate}

Since these are highly coupled non-linear differential equations, they must be simplified to obtain analytical solutions. Figure.~(\ref{Flowchart1}) contains the procedure we have adopted to simplify the modified Einstein tensor components \eqref{eq:Secord-ModEqs} such that we separate the $\chi$ dependent and independent terms. More specifically, we explicitly show that the procedure we adopt can reduce the fourth-order differential equation to a product of two second-order differential equations. We now provide detailed steps to derive the the solutions:
\begin{figure}[!h]
\centering
\includegraphics[width=0.6\textwidth]{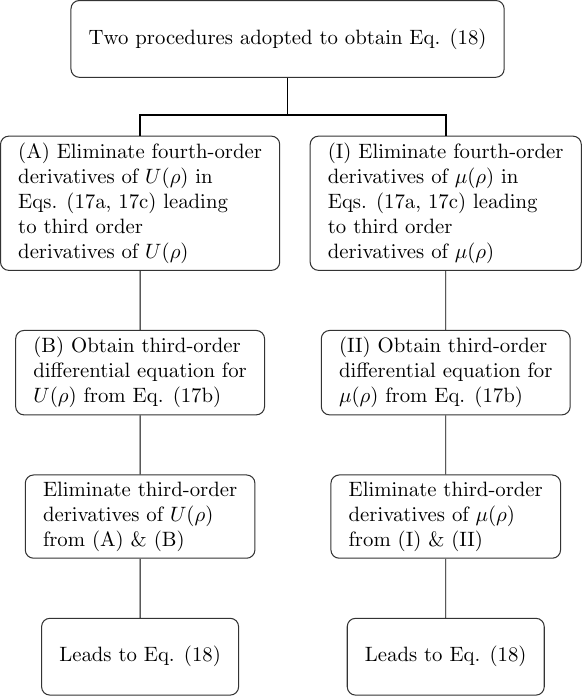}
\caption{Procedures to obtain \eqref{EQM2ndorderForm}}
\label{Flowchart1}
\end{figure}
\begin{steps} 
\item We eliminate the fourth-order derivatives of $U(\rho)$ from $T_1$ \eqref{eq:Secord-ModEqsa} and $T_3$ \eqref{eq:Secord-ModEqsc}. It is possible to eliminate either $U(\rho)$ or $\mu(\rho)$ to obtain a third-order differential equation in $\mu(\rho)$ or $U(\rho)$. As shown in Fig. 1, the two procedures lead to identical equation.
\item Using the third-order differential equation of $U(\rho)$ (or $\mu(\rho)$ from Step 1 and the third-order differential equation $T_2$ \eqref{eq:Secord-ModEqsb}, we obtain the following equation:
%
\begin{equation}
\begin{aligned}\label{EQM2ndorderForm}
    \frac{W_{0}(\rho)}{4\,\alpha_{1}^2\,\rho^5\,U(\rho)\left[\Phi(\rho) -2\right]^2[\Phi(\rho) + 4]^2}\,\chi^2 + \frac{W_{1}(\rho)\,W_{2}(\rho)}{2\,\alpha_{1}^2\,\rho^3\,U(\rho)\left[\Phi(\rho) -2\right][\Phi(\rho) + 4]} = 0
\end{aligned}
\end{equation}

%
where, $\Phi(\rho) = \rho[\mu'(\rho) + [\ln\,U(\rho)]']$. $W_1(\rho)$ and $W_2(\rho)$ are independent of $V(\rho)$, while $W_0$ is dependent of $V(\rho)$. Note that $W_0(\rho), W_1(\rho)$ and $W_2(\rho)$ are second-order differentials. 
The above expression allows us to evaluate $U(\rho)$ and $\mu(\rho)$ by solving $W_1(\rho)$ or $W_2(\rho)$ without the knowledge of $V(\rho)$. This is crucial in obtaining SRBH solutions in $f(R)$.
\item Since $W_1(\rho)$ and $W_2(\rho)$ are independent of $V(\rho)$, we can obtain the exact form of $U(\rho)$ and $\mu(\rho)$ by setting $\chi = 0$ in the above expression. This leads to the following conditions:
\[
W_1(\rho) = 0 \quad\mbox{or}\quad W_2(\rho) = 0 \, .
\]
We can choose either one of these conditions to obtain the exact form of $U(\rho)$ and $\mu(\rho)$. Note that the form of $W_1(\rho)$ and $W_2(\rho)$ are different; hence, the solutions one obtains will be different.

We obtain the exact form of $U(\rho)$ and $\mu(\rho)$ in two steps. First, we set 
$\mu(\rho) = 0$ in $W_2(\rho) = 0$. This leads to a simplified differential equation for $U(\rho)$ solving, which leads to:
\begin{equation}
U(\rho) = 1 - \frac{C_1}{\rho} + \frac{C_0}{\rho^2} + \frac{\rho^2}{12 \kappa^2} 
\label{eq:SolofU-V}    
\end{equation}
where $C_1$ and $C_0$ are integration constants. Details of the derivation are provided in \ref{App:B}.
\item Having obtained $U(\rho)$, we now substitute $U(\rho)$ in Eq.~\eqref{EQM2ndorderForm} and solve the resultant differential equation for $\mu(\rho)$.  For $C_0 = 0$, we obtain the following closed form expression for $\mu(\rho)$:
\small{
\begin{align}
\label{eq:musolfin}
\frac{\mu(\rho)}{2} = 
\ln \left[1 + {N_1} \left(\int \rho^{-1/2}\left(\rho^3 + 12\,\rho\,\kappa^2 - 12\,C_{1}\,\kappa^2\right)^{-3/2} \,d\rho\right) \right] ,
\end{align}
}
where $N_1$ is an integration constant. $N_1 < 0$ ensures that $\exp[\mu(\rho)]$ is positive definite for all values of $\rho$. For $C_1 \neq 0$, it is possible to obtain a closed-form expression; however, they do not provide any physical insight. Hence, in the rest of the analysis, we obtain slowly-rotating solutions for $C_0 = 0$.

\item Having obtained $U(\rho)$ and $\mu(\rho)$, we 
now need to determine $V(\rho)$. To do this, we substitute $U(\rho)$ and $\mu(\rho)$ (obtained in Step 3 and 4 above) in Eq.~\eqref{eq:Secord-ModEqs}. Demanding that all the components of the modified Einstein tensor \eqref{eq:Secord-ModEqs} decays asymptotically leads to the following form:
\begin{equation}
V(\rho) = \sum_{j=0}^{\infty}\frac{C_{j + 2}}{\rho^{j+2}}
\label{eq:Solof_V}    
\end{equation}
where $C_{j + 2}$ are constants to be determined by physical conditions. Note that $j = 1$ results in slowly-rotating Kerr-like metric~\cite{2016-Ayzenberg_Dimitry_Yunes-CQG}. In the rest of this analysis, we consider the following form of $V(\rho)$: 
\[
V(\rho) = C_3 \, \rho^{-3} \, .
\]
In \ref{App:C}, we compare the kinematical properties of the SRBH with GR for $V(\rho) = C_2 \, \rho^{-2}$.

\end{steps}
This is the crucial result of this work, regarding which we would like to discuss the following points: First, to our knowledge, this is the first time a non-trivial rotating BH solution has been obtained in any $f(R)$ gravity model. Note that this is an approximate solution in the slowly-rotating limit $\chi \ll 1$. As shown in \ref{App:B}, substituting the above forms of $U(\rho), V(\rho)$ and $\mu(\rho)$ in Eq.~\eqref{EQM2ndorderForm} leads to a vanishing of $\chi^2$ dependent terms asymptotically.

Second, according to GR, Kerr (vacuum) and Kerr-Newman (with charge) are the final states of gravitational collapse. BHs with scalar or other kinds of hair, i.e., outer space-time characterized by more than three parameters, are impossible in GR. However, in $f(R)$ gravity, the generalized Bianchi identity \eqref{GeneralizedBianchi} provides a non-trivial structure for the Ricci scalar as a function of $\rho$ leading to an infinite set of static slowly rotating black hole (SRBH) solutions for $f(R)$ gravity. Besides, there are two branches ($W_1(\rho) = 0$ or $W_2(\rho) = 0$) of SRBH solutions, and a collapse of a star might lead to a BH in either one of these branches. Note that the constants $C_{2}, C_3, C_4, \cdots$ are arbitrary, and these lead to the infinite choice of slowly-rotating black-hole solutions in $f(R)$. Thus, as mentioned above, the SRBH solutions we have obtained depend on the coupling parameters $\alpha_{0}$ and $\alpha_{1}$ via these constants $C_{2}, C_3, C_4, \cdots$. In GR, the Plebański–Demiański black hole solution is an example of rotating BH in asymptotic non-flat space-times~\cite{1976-Plebanski}. Specifically, Plebański–Demiański black hole solution has six free parameters. In our case, we have three unknown functions --- $U(\rho), V(\rho)$ and $\mu(\rho)$. $\mu(\rho)$ has one unknown parameter, $U(\rho)$ has maximum 3 parameters, and $V(\rho)$ has infinite parameters. Thus, we see the slowly rotating BH solutions in  $f(R)$ model has infinite number of unknown parameters. These parameters are only valid if $\alpha_0 \neq 0$. Hence, the results are valid only for the non-zero cosmological constant. Hence, our results indicate that the no-hair theorem for modified gravity theories merits extending to include the coupling constants.
In the specific $f(R)$ we have considered, we need to include the two coupling parameters to uniquely describe a rotating black hole.

Third, for the $f(R)$ model \eqref{f(R)2ndorder}, two branches of solutions exist: $W_1(\rho) = 0$ or $W_2(\rho) = 0$. By setting $W_2(\rho) = 0$, we have explicitly constructed two non-trivial slowly-rotating solutions. Specifically, (i) $U(\rho)\neq 0$, and $\mu(\rho)=0$ and  (ii) $U(\rho)\neq 0$, and $\mu(\rho)\neq 0$. However, the above choices are among the \textit{infinite} choices of the metric components.
%
In the $\chi = 0$ limit, we can identify $C_1$ with the mass of the BH, and hence, we have $C_1 = 2$. However, we can not determine the exact value of other coefficients ($C_3, C_4$), except that they have to be positive definite.

Fourth, we want to contrast our approach to the other approaches used in the literature to obtain exact solution~\cite{Cikintoglu:2017jfh,Molano:2020ocd}. The approach used in Refs.~\cite{Cikintoglu:2017jfh,Molano:2020ocd} complements our current work. These authors employ numerical methods to investigate a boundary layer near the surface of a star, where higher-order terms in the gravitational equations become negligible. Beyond this layer, the well-established SdS (Schwarzschild-de Sitter) solution is applied asymptotically to describe the gravitational field. Meanwhile, within the boundary layer, the numerical technique is used to derive an interior solution that aligns with the external solution across the boundary. However, a limitation of this approach is evident as it fails to produce a unique solution. This is due to the reliance on the Ricci scalar value at the star's surface to determine the resolution of the interior solution. This lack of a universal solution stems from the absence of a theorem ensuring consistency among theories formulated using this method. 

Fifth, to obtain exact solutions in $f(R)$, one usually performs a conformal transformation~\cite{2012-Sotiriou-Faraoni-PRL,2015-Canate.etal-CQG,2018-Sultana.Kazanas-GRG,2018-Calza.etal-EPJC,2021-Nashed.Nojiri-PRD}. Here, we have not made any approximation or performed a conformal transformation to obtain exact solutions. Our analysis and results are valid even if the conformal transformations to the Einstein frame are not well-defined. 
Under conformal transformations ($\tilde{g}_{\mu\nu} =  F(R) \, g_{\mu\nu}$), the action~ Eq.~\eqref{f(R)2ndorder} transforms to~\cite{2010-Sotiriou_Faraoni-RMP,2010-DeFelice.Tsujikawa-LivingRev.Rel.,2007-Woodard-LNotePhy}:
\begin{equation}
S=\int d^4x\sqrt{-\tilde{g}}\left[\frac{1}{2\kappa^2}\tilde{R}-\frac{1}{2}\partial^{\alpha} {\varphi}\partial_{\alpha}{\varphi}-V({\varphi})\right] \, ,
\end{equation}
where, ${\varphi},$ and $V(\varphi)$ are given by:
\begin{eqnarray}
\sqrt{\frac{2 \kappa^2}{3}}  \varphi &=&  
  (p - 1) \ln (\alpha_0 + \alpha_1 R) + \ln(\alpha_1 p) \\
V(\varphi) &=& \frac{1}{2 \kappa^2 p^2 \alpha_1} 
\frac{(p - 1) \alpha_1 R - \alpha_0}{(\alpha_0 + \alpha_1 R)^{p -1} } 
\end{eqnarray}
In the case of spherically symmetric metric the solutions satisfy the condition $R = -\alpha_0/\alpha_1$, hence,  $V(\varphi)$ diverges and the theory is ill-defined in the Einstein frame. Thus, such $f(R)$ models do not have an equivalent description in the Einstein frame. It is important to note that other authors have pointed out the non-equivalence of the Jordan and Einstein frames in other $f(R)$ models~\cite{2007-Briscese_Nojiri_Odintsov-PLB,2006-Capozziello_Nojiri_Odintsov-PLB,2010-Capozziello_MartinMosruno_Rubano-PLB,2015-Cognola-PRD}.

However, for the slowly rotating black hole solutions to the quadratic order in $\chi$. Using the metric components from Eqs.~\eqref{eq:SolofU-V} - \eqref{eq:Solof_V}, we get:
\begin{align}
   f(R)&=\left( \alpha_0+\alpha_1\left\{\frac{-12\alpha_{0}^{3} \rho^{6}+288 \alpha_{0}^{2}\alpha_1 \rho^3(\rho-C_1)-1728\alpha_{1}^{2}\alpha_0(\rho\,[\rho-2C_1]+C_1^{2})}{12\alpha_1(-\alpha_0\rho^3+12C_1\alpha_1-12\alpha_0\rho)^2}+\chi^{2}(\cdots)\right\} \right)^2\label{refref_R}\\
   f'(R)&=2\,\alpha_1\left( \alpha_0+\alpha_1\left\{\frac{-12\alpha_{0}^{3} \rho^{6}+288 \alpha_{0}^{2}\alpha_1 \rho^3(\rho-C_1)-1728\alpha_{1}^{2}\alpha_0(\rho\,[\rho-2C_1]+C_1^{2})}{12\alpha_1(-\alpha_0\rho^3+12C_1\alpha_1-12\alpha_0\rho)^2}+\chi^{2}(\cdots)\right\} \right)\label{refref_R_der}
\end{align}
Note that $ f''(R) = 2 \alpha_1^2 > 0$ everywhere in the space-time. 
At spatial infinity ($\rho \rightarrow\infty$)
\[
R \rightarrow -\frac{\alpha_{0}}{\alpha_{1}},~f(R) \rightarrow 0, ~
 f'(R) \rightarrow 0
\]
This hold true irrespective of $\mu(\rho) = 0$ and $\mu(\rho) \neq 0$. We see that $f'(R)$ vanishes only at asymptotic infinity. As mentioned in Sec. 7.4 of Ref.
\cite{2010-DeFelice.Tsujikawa-LivingRev.Rel.}, the condition $f'(R) > 0$ will be satisfied, implying that the conformal transformation is well-defined.

Lastly, to understand the properties of the space-time, we can evaluate the curvature scalars such as $R_{\mu \nu\sigma\rho}R^{\mu \nu\sigma\rho}$, $R_{\mu \nu\sigma\rho}R^{\mu \sigma}R^{\nu \rho}$, and $R_{\mu \nu}R^{\mu \nu}$, and expanded these curvature scalars to the second order in the rotational parameter~$\chi$  for the metric components:
 
\begin{align*}
   U(\rho) =& 1 - \frac{C_1}{\rho} + \frac{\rho^2}{12 \kappa^2}, \quad
   \mu(\rho)= 0 ,\quad
    V(\rho) = \frac{C_3}{ \rho^{3}}, 
\end{align*}
and  we obtain the following curvature scalars:
\begin{align*}
    R_{\mu \nu\sigma\rho}R^{\mu \nu\sigma\rho} &=\mathcal{S}^{(0)}_{1}\left(\frac{1}{\rho^6}\right)+\chi^2~\mathcal{S}^{(II)}_{1}\left(\frac{1}{\rho^8}\right)\\\
    R_{\mu \nu\sigma\rho}R^{\mu \sigma}R^{\nu\rho} &=\textit{Constant}+\chi^2~\mathcal{S}^{(II)}_{2}\left(\frac{1}{\rho^2}\right)\\
    R_{\mu \nu}R^{\mu \nu} &=\textit{Constant}+\chi^2~\mathcal{S}^{(II)}_{3}\left(\frac{1}{\rho^2}\right).
\end{align*}
where $\mathcal{S}^{(0)}_{1}, \mathcal{S}^{II)}_{1}, \mathcal{S}^{(II)}_{2}, \mathcal{S}^{(II)}_{3}$ are functions~$\rho$. We can compare the above expressions with the curvature scalars of the slowly rotating Kerr black hole solution (in GR):
\begin{align*}
    R_{\mu \nu\sigma\rho}R^{\mu \nu\sigma\rho} &=\mathcal{P}^{(0)}_{1}\left(\frac{1}{\rho^6}\right)+\chi^2~\mathcal{P}^{(II)}_{1}\left(\frac{1}{\rho^8}\right)\\\
    R_{\mu \nu\sigma\rho}R^{\mu \sigma}R^{\nu\rho} &=\mathcal{O}(\chi^4)\approx \chi^4~\mathcal{P}^{(IV)}_{2}\left(\frac{1}{\rho^7}\right)\\
    R_{\mu \nu}R^{\mu \nu} &=\mathcal{O}(\chi^4)\approx \chi^4~\mathcal{P}^{(IV)}_{3}\left(\frac{1}{\rho^{10}}\right)
\end{align*}
where $\mathcal{P}^{(0)}_{1}, \mathcal{P}^{II)}_{1}, \mathcal{P}^{(IV)}_{2}, \mathcal{S}^{(IV)}_{3}$ are functions~$\rho$. 
Comparing the above two sets of expressions, it is clear that all the curvature scalars of the new slowly-rotating solutions in $f(R)$ have singularity at $\rho = 0$ and are finite everywhere, similar to the slowly rotating Kerr black hole solution.

In the next section, we discuss the key properties of these solutions and compare with Kerr and SR Kerr solution.

\section{Properties of the SRBH solutions} 
\label{sec:Properties}

\begin{figure*}[!htb]
\begin{minipage}[b]{.482\textwidth}
	\centering
    \subcaptionbox{ \label{fig:horizon} }{
\includegraphics[width=\columnwidth]{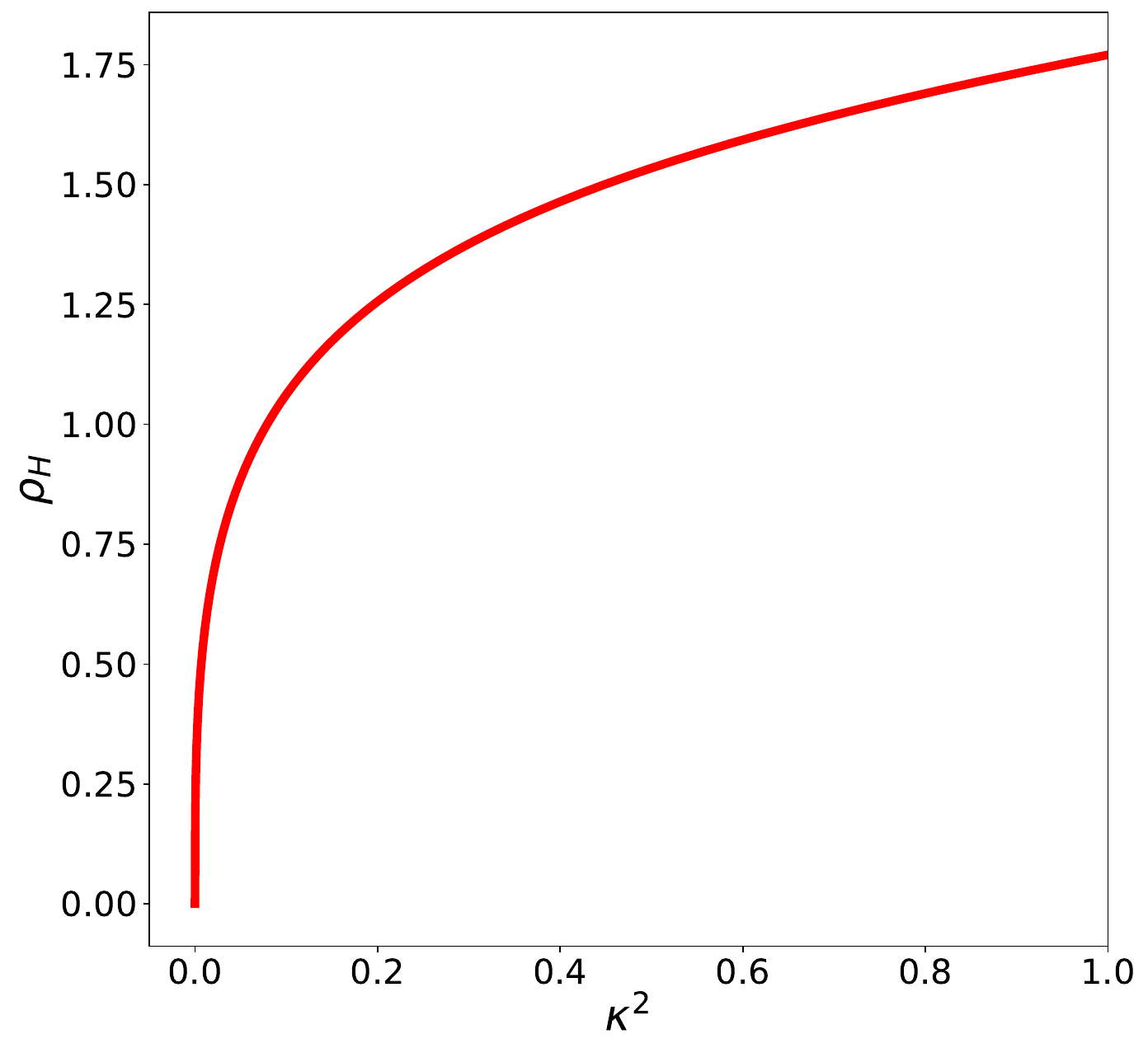} }

\end{minipage}\hfill
\begin{minipage}[b]{.45\textwidth}
\centering
    \subcaptionbox{ \label{fig:ergo}}{
\includegraphics[width=\columnwidth]{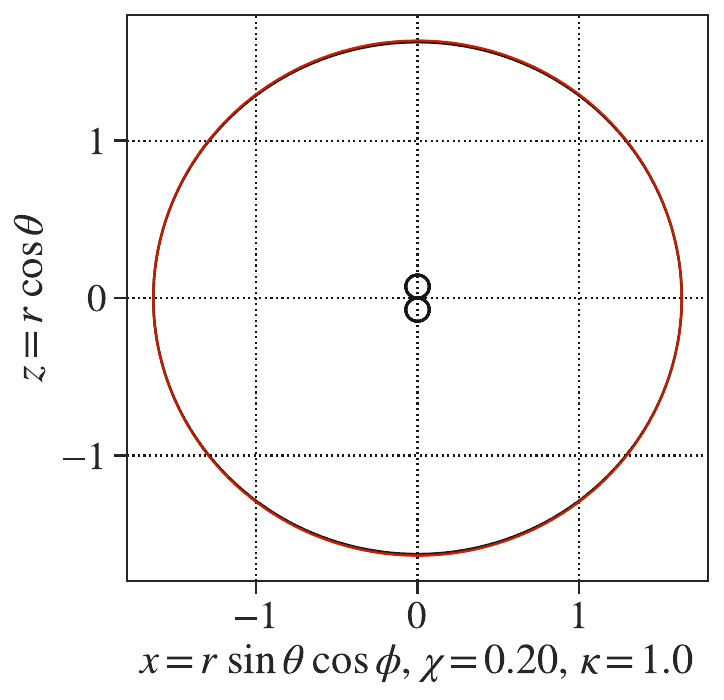}}
\end{minipage} 

\caption{(a) Plot of $\rho_H$ as a function of $\kappa^2$ for $\chi = 0.1$ and $C_3 =1$. (b) Plot of ergosphere for SRBH in $f(R)$ for $\chi = 0.2$ and $\kappa^2 = 1$.}
\label{sym}
\end{figure*}

First, these solutions correspond to a BH with an identical event horizon. The event horizon is determined by the condition $g^{\rho \rho} = 0$. For our metric, we have three solutions: one real and two imaginary. The horizon $\rho_H$ is given by:
\begin{equation}
\label{eq:horizon}
\rho_H = \frac{H^{2/3}\,\left(2\,\kappa^2\right)^{1/3} - 2^{5/3}\,\kappa^{4/3}}{H^{1/3}}\,;~~
H = 2\,\sqrt{9 + 4\,\kappa^2} + 6 \, .
\end{equation}
Note that this matches with the slowly-rotating limit of Kerr BH ($\rho_{\rm SR} = 2$) for small values of $\kappa^2$~\cite{2016-Ayzenberg_Dimitry_Yunes-CQG}. To see this, doing a series expansion of Eq.~\eqref{eq:horizon} about small $\kappa^2$, we have:
$$
\rho_H \approx (24)^{1/3}\,\kappa^{2/3} 
$$
Thus, for $\kappa^2 = 0.33$, the above solution matches the SR Kerr solution in GR. As seen from Fig.~\eqref{fig:horizon}, the horizon radius can be larger or smaller than the SR Kerr depending on the value of $\kappa^2$; the larger the value of $\kappa^2$, the larger the horizon radius compared to GR. This will be crucial when we discuss the kinematical properties of space-time. Since the above condition $g^{\rho \rho} = 0$ leads to only one real solution, the slowly rotating BH solution we obtained do not have any cosmological horizon. This is similar to the behavior of slowly rotating Kerr-dS solution in GR.

As mentioned above,  we consider the following form of $V(\rho)$ for the rest of the analysis in this section: 
$$
V(\rho) = C_3 \, \rho^{-3} \, .
$$
In order to confirm that the solutions indeed correspond to BH, we evaluate the Kretschmann scalar: 
\begin{align}
K_{\mu(\rho) = 0} &= \frac{1}{6\,\kappa^2} + \frac{48}{\rho^6} + \chi^2 \, 
{\cal F}_{0}\left(\frac{1}{\rho}\right)
 \\
K_{\mu(\rho) \neq 0} &= \frac{1}{6\,\kappa^2} +  {\cal F}_{1}\left(\frac{1}{\rho}\right) 
+ \chi^2 {\cal F}_{2}\left(\frac{1}{\rho}\right)
\end{align}
where ${\cal F}_0, {\cal F}_1, {\cal F}_2$
 are functions of $\rho$. All these three functions diverge at the origin and vanish at infinity. 

In order to confirm that these two solutions correspond to rotating BHs, we turn our attention to the ergosphere. 
The radius of the ergosphere ($g_{\tau\,\tau} =0$) is given by:
\begin{align}
    \frac{1}{12\,\kappa^2}\rho^5 + \rho^3 - 2 \rho^2 +C_{3}\,\chi^2\, \cos^2(\theta) = 0
\end{align}
Note that $C_3 > 0$. Fig.~\eqref{fig:ergo} is the $x-z$ plot of the ergosphere for $\chi = 0.2$ and $\kappa^2 = 1$. The red-curve corresponds to the event-horizon, while the two black curves correspond to the ergosphere. Note that the variation due to $\chi^2\,\cos{\theta}$ is visible in the inner part of the ergoregion. As expected the event horizon is very close to the ergosphere for SRBHs.

Having confirmed that $f(R)$ model \eqref{eq:fRact-Jordan}, an attentive reader might wonder how to distinguish this in future observations. Comparing the GW signatures of these solutions requires one to solve the perturbed modified equations, like in the case of Chern-Simons~\cite{2021-Srivastava.etal-PRD} and will be discussed in future work. Here we discuss the kinematical properties, including conserved quantities, that can be used in the ngEHT observations~\cite{2022-Tiede.ngEHT-Galaxies}. \\[5pt]

%

\subsection{Kinematical properties of the SRBH space-times}

In this subsection, we obtain the kinematical properties of the SRBH space-times and compare them with Kerr solution in GR. In \ref{App:D}, we compare the kinematical properties of the SRBH space-times with asymptotically non-flat rotating black hole solution in GR~\cite{2009-Lu_Pope-IOP,1973-Demianski-Acta_Astro,2005-Gibbons-JGP}.

Like Kerr, the SRBH space-times \eqref{Slow_Kerr} possess two --- time-like and azimuthal --- Killing vectors, leading to the conservation of the specific energy $(E)$ and the axial component of the specific angular momentum $(L_{z})$. $E$ and $L_z$ can be rewritten in terms of the 4-velocity $(u^{\mu})$ of the time-like particle: 
\begin{equation}
E =-\left(g_{\tau \tau} u^\tau+g_{\tau \phi} u^\phi\right); \quad L_z =g_{\phi \tau} u^\tau+g_{\phi \phi} u^\phi \, .
\end{equation}
\begin{figure*}[!htb]
\begin{minipage}[b]{.5\textwidth}
	\centering
    \subcaptionbox{}{
\includegraphics[width=\columnwidth]{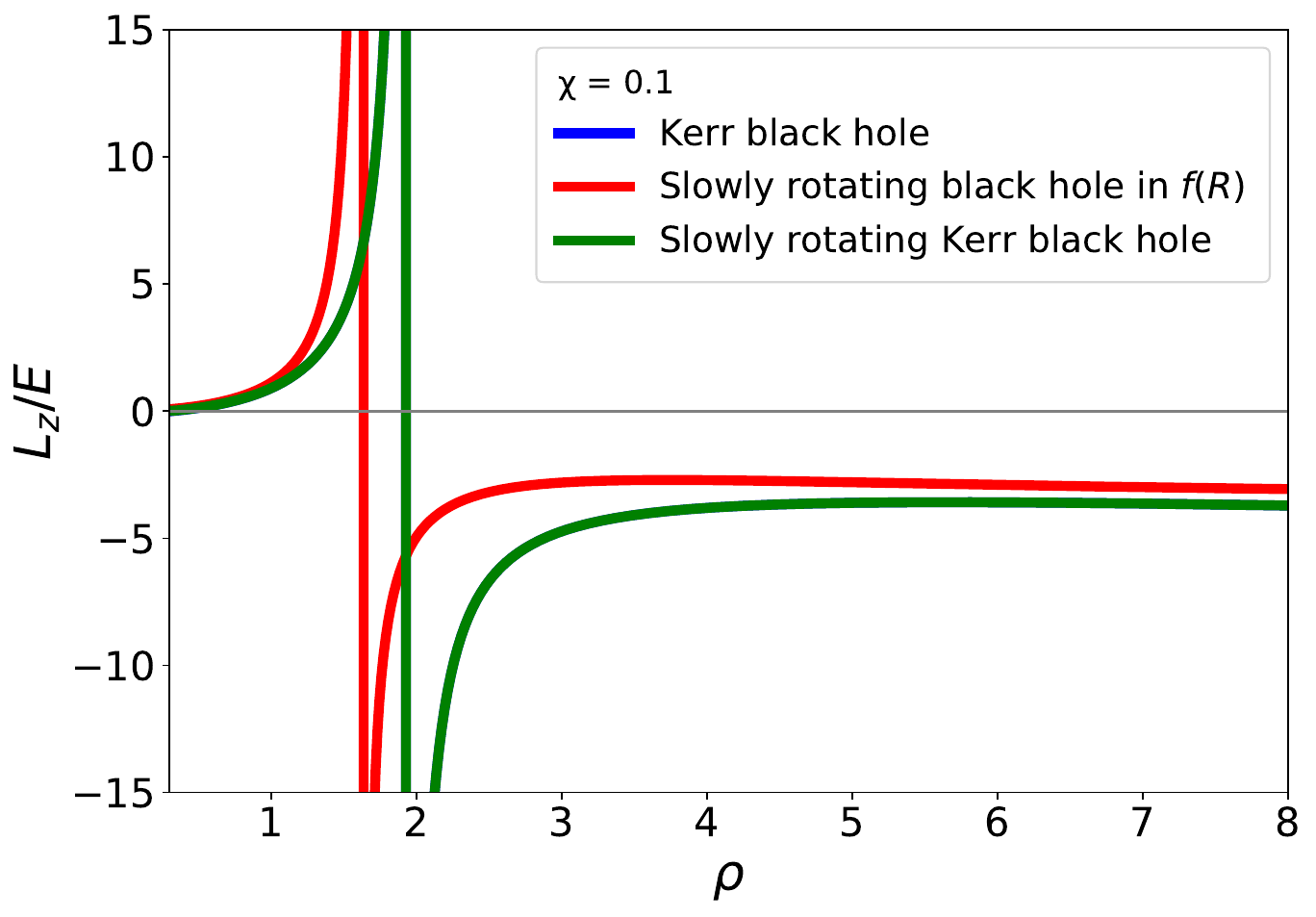}  }
\label{fig:LzEchi+ve}
\end{minipage}\hfill
\begin{minipage}[b]{.5\textwidth}
	\centering
    \subcaptionbox{}{
\includegraphics[width=\columnwidth]{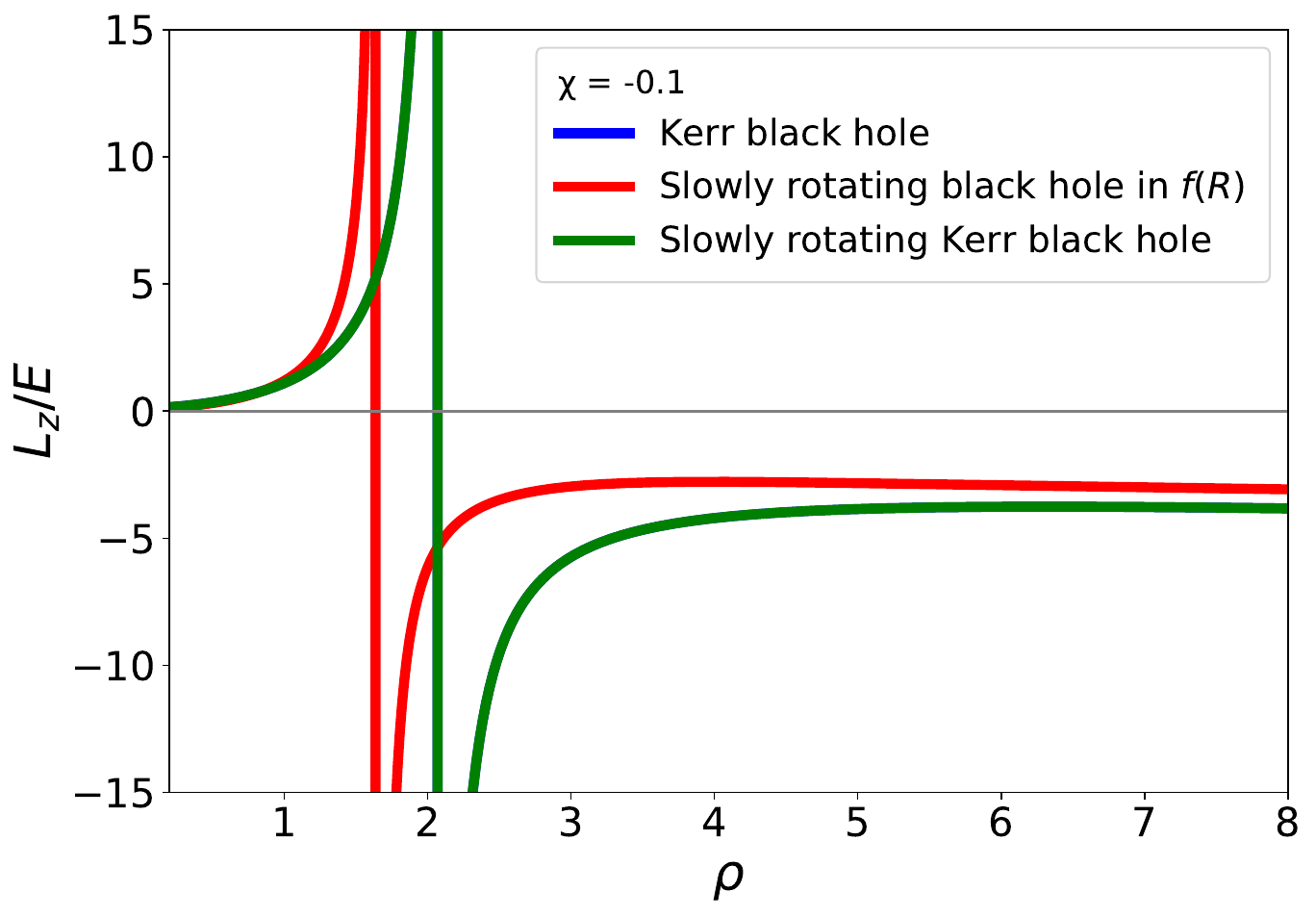}}
\label{fig:LzEchi-ve}
\end{minipage} 
\caption{Plot of $L_z/E$ of a test particle in equatorial circular orbits as a function of $\rho$ for $\kappa^2 =1$ and $C_1 =2$. We also plot for SR Kerr solution and Kerr solution. (a) For  $\chi = 0.1$. (b) For  $\chi = - 0.1$.}
\label{fig:LzE}
\end{figure*}

Since the geometry of the SR metric is axisymmetric, the orbital paths of objects about these black holes are often complex. To highlight the difference between the SRBH in $f(R)$ and GR, we consider the orbits in the equatorial plane ($\theta = \pi/2$) of the BH.
Focusing on the equatorial circular orbits ($\dot{r} = 0$) around the BH satisfying the condition $u_{\mu} u^{\mu} = -1$, we have:
\begin{equation}
\begin{aligned}
\label{eq:defELz}
E&=-\frac{g_{\tau\tau}+\Omega \, g_{\tau\phi }}{\sqrt{-g_{\tau\tau}-2\Omega \, g_{\tau\phi}-\Omega^2g_{\phi\phi} } }; \\
L_{z} &=\frac{g_{\tau\phi}+\Omega \, g_{\phi\phi }}{\sqrt{-g_{\tau\tau}-2\Omega \, g_{\tau\phi}-\Omega^2g_{\phi\phi} } } \, ,
\end{aligned}
\end{equation}

where the angular velocity ($\Omega$) is
\begin{align}
\Omega&=\frac{-\partial_{\rho}g_{\tau \phi} +\sqrt{\left( \partial_{\rho}g_{\tau \phi}\right)^2-\partial_{\rho}g_{\tau\tau}\partial_{\rho}g_{\phi \phi}} }{\partial_{\rho}g_{\phi \phi} } \, .
\end{align}
Fig. \eqref{fig:LzE} contains the plot of $L_z/E$ by evaluating up to the second order in $\chi$ for $\mu(\rho) = 0$. Like in the previous plots, we have compared the results of $L_z/E$ for slowly-rotating $f(R)$ BH space-times with Kerr and slowly-rotating Kerr\footnote{Note that the characteristics of the Kerr BH are similar to those of the slowly rotating Kerr BH. Hence, in Fig. \eqref{fig:LzE}, we will not be able to distinguish the characteristic between the Kerr and the slowly rotating Kerr BH. In other words, a slowly rotating Kerr follows the same curve as the Kerr metric.}. Note that $L_z/E$ is zero at the horizon for all the three BH space-times and diverges at the horizon. Since $L_z/E$ is related to the impact parameter, from the figure, we infer that the impact parameter of the BHs in $f(R)$ is smaller than that of Kerr. This means that the inner-most stable circular orbit for BHs in $f(R)$ is smaller; hence, the shadow radius might also be smaller. Thus, ngEHT can potentially be used to constrain $\kappa^2$. This is currently under investigation. 

\begin{figure*}[!htb]
\begin{minipage}[b]{.5\textwidth}
	\centering
    \subcaptionbox{  \label{fig:EffV}  }{
\includegraphics[width=\columnwidth]{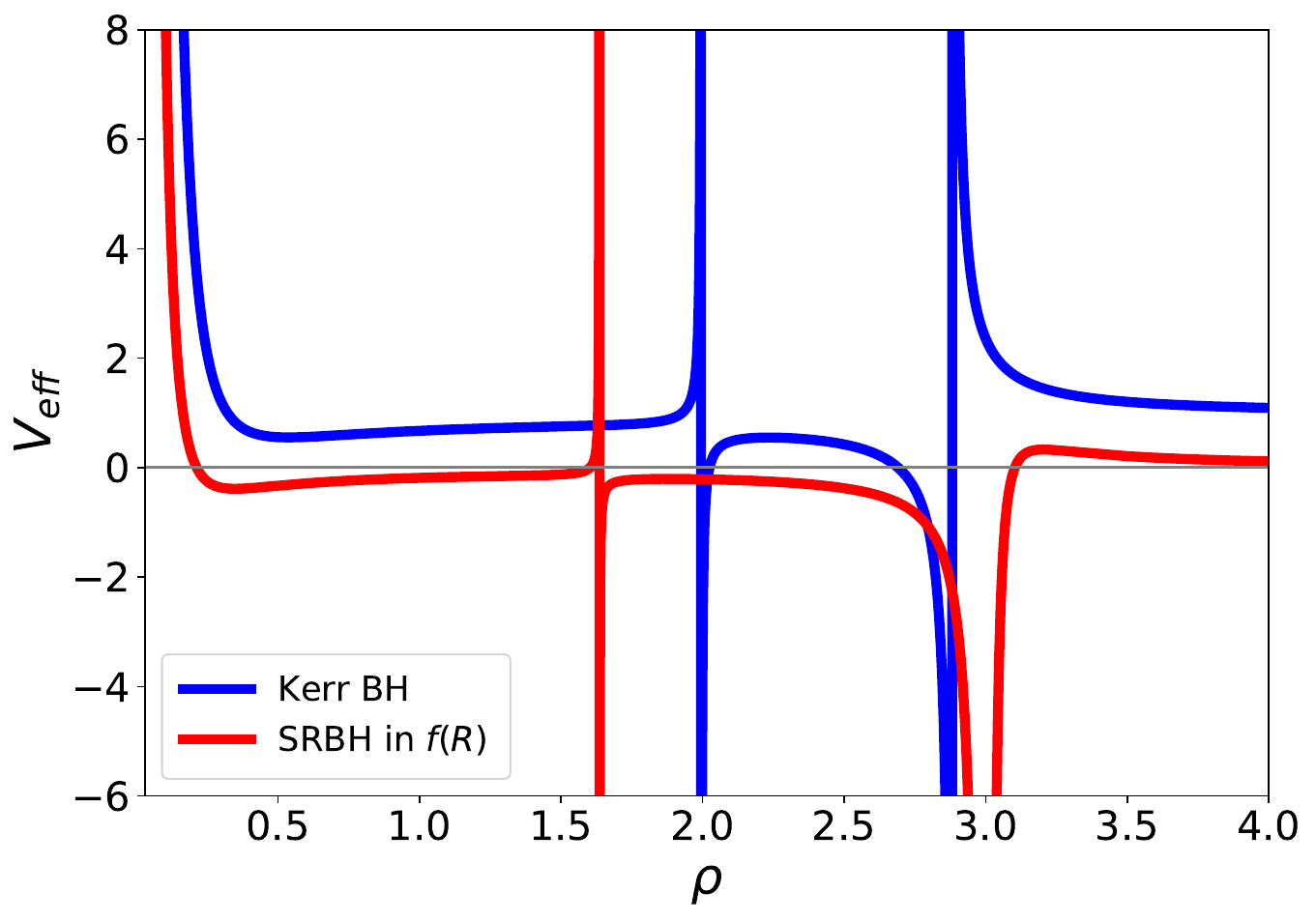}  }

\end{minipage}\hfill
\begin{minipage}[b]{.5\textwidth}
	\centering
    \subcaptionbox{ \label{fig:zamo}  }{
\includegraphics[width=\columnwidth]{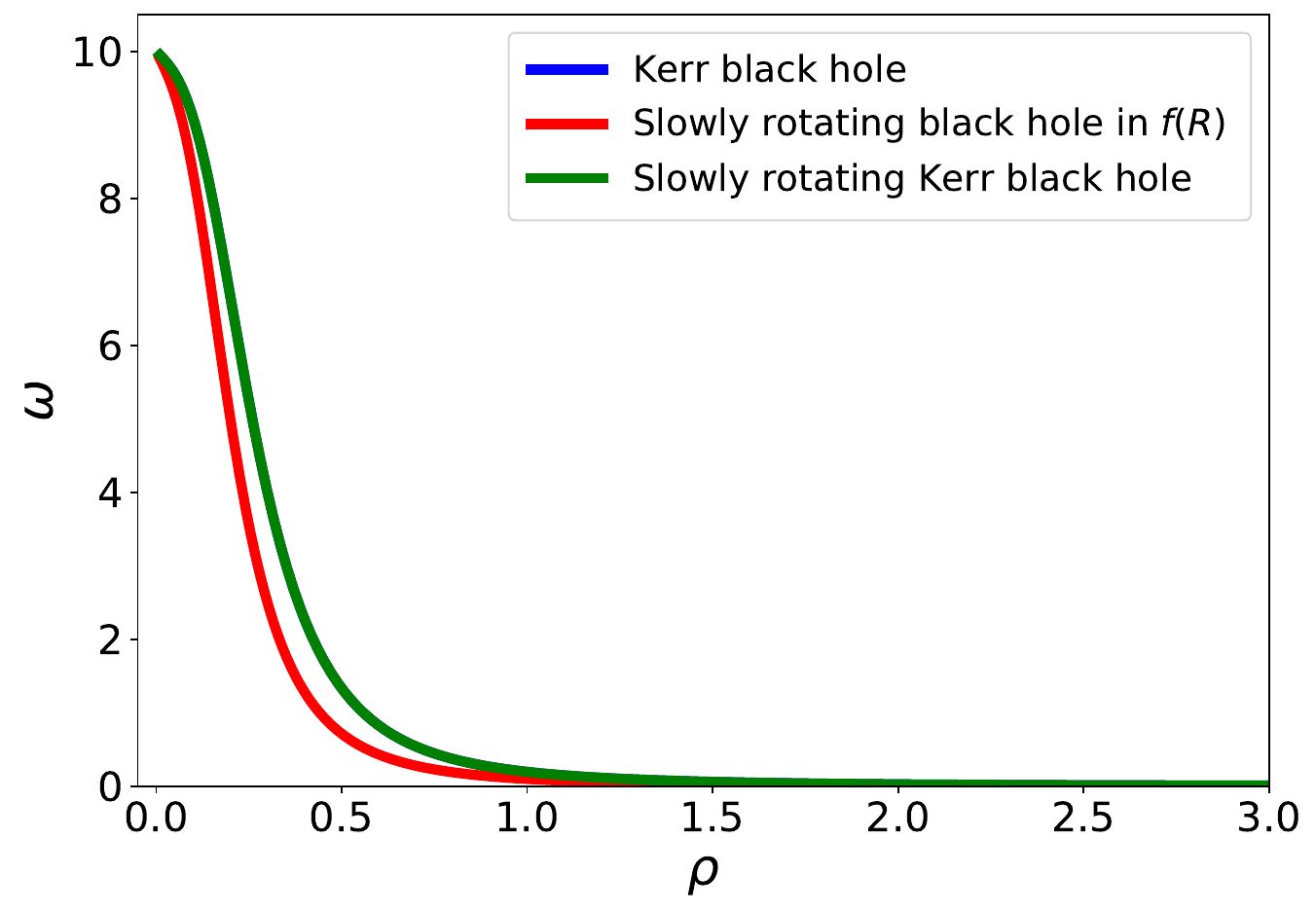}   }

\end{minipage} 
\caption{(a) Plot of $V_{\rm eff}$ of a test particle in equatorial plane as a function of $\rho$. We set $C_1 = 2$.
(b) Plot of the equatorial plane ZAMO's angular velocity as a function of $\rho$. For line-element  \eqref{Slow_Kerr}, we have set $\kappa^2 = 1, \chi = 0.1$ and $C_3 =1$. We also plot for SR Kerr and
exact Kerr.}
\label{fig:VeffZamo}
\end{figure*}

Using Eq.~\eqref{eq:defELz}, the radial equation on the equatorial plane is~\cite{1973-Misner.etal-Book,1985-Chandrasekhar-Book,2012-Bamba.etal-Astrophys.SpaceSci.}:
\begin{equation}
\frac{E^2-1}{2}=\frac{1}{2}\left(\frac{d \rho}{d \tau}\right)^2+V_{\mathrm{eff}}(\rho, E, L_z)
\end{equation}
where $V_{\rm eff}$ is the effective potential of the test particle given by:
\begin{align}
    V_{\rm eff}&=\frac{E^2 g_{\phi \phi}+2EL_{z}g_{\tau \phi}+L^{2}_{z} g_{\tau \tau}}{g^{2}_{\tau \phi}-g_{\tau \tau}g_{\phi \phi}}-1
\end{align}
Fig.~\eqref{fig:EffV} contains the plot of $V_{\rm eff}$ of a test particle in the equatorial plane for $\mu(\rho) = 0$.
Like in the previous plots, we have compared the results of $V_{\rm eff}$ for SRBH in $f(R)$ with Kerr and SR Kerr. It is known that $V_{\rm eff} = 0$ corresponds to the circular orbits in 
the equatorial plane. These plots show that the circular orbits in the SRBH in $f(R)$ are smaller than Kerr.  Note 
that $V_{\rm eff}$ diverges near the horizon.
Again, we want to point out that for $\kappa^2 = 1$, the horizon radius of the slowly-rotating $f(R)$ black hole is larger than GR.
The divergence near $\rho = 3$ corresponds to the existence of 
ISCO~\cite{1985-Chandrasekhar-Book,2012-Bamba.etal-Astrophys.SpaceSci.}. This needs detailed investigation which is beyond the scope of this work.
%
%

%
%

Let us now consider an observer moving with four-velocity $u^{\mu}$ in the stationary axisymmetric space-time \eqref{Slow_Kerr}. A special class of observers (ZAMO) has the property that their angular momentum vanishes. More specifically, the angular velocity $\omega$ vanishes for a ZAMO at infinity, but in the general case, it is nonzero and position-dependent. ZAMOs are defined by the condition~\cite{1973-Misner.etal-Book,1985-Chandrasekhar-Book,2012-Bamba.etal-Astrophys.SpaceSci.}:
\begin{equation}
g_{\tau \phi} \dot{\tau} 
+ g_{\phi \phi} \dot{\phi}=0    
\end{equation}
For the line-element \eqref{Slow_Kerr}, the ZAMO’s angular velocity is
\begin{align}
\omega &=-\frac{g_{\tau\phi}}{g_{\phi \phi}} = \frac{\chi\,C_{3}}{\rho^3+\chi^2\left(C_3\,\sin^2(\theta) 
+\rho\right)} \, . 
\end{align}
Fig. \eqref{fig:zamo} contains ZAMO's angular velocity for the SR $f(R)$ space-time and compares it with Kerr and slowly-rotating Kerr for the equatorial plane ($\theta = \pi/2$). From the figure, we infer that the ZAMO's angular velocity for an SRBH in $f(R)$ is always smaller than that of Kerr. This is intriguing because, for $\kappa^2 = 1$, the horizon radius of the slowly-rotating $f(R)$ black hole is larger than GR.  

\section{Conclusions and Discussions} 
\label{sec:conc}

For a class of $f(R)$ models, we have obtained asymptotically non-flat, SRBH solutions up to second-order in the rotational parameter $\chi$. Specifically, we obtained mutiple BH solutions with the event horizon at the same point with different asymptotic features. Besides there are two branches ($W_1(\rho) = 0$ or $W_2(\rho) = 0$) of SRBH solutions and a collapse of a star might lead to a BH in either one of these branches. 

{As discussed in detail in \ref{sec:refs}, in GR, no-hair theorems concern mostly black hole solutions with flat asymptotics. It has been shown that the \emph{standard} no-hair theorem in GR fails in some cases, like Einstein–Yang–Mills theory, nonlinear electrodynamics, Dilaton gravity, Einste-Gauss-Bonnet-Dilaton gravity and $R^2$, among others~\cite{1986-Hugh_Ian-PLB,1989-Volkov-JETP,1993-Greene-PRD,1991-Markus_Serge_NorbertStraumann-PLB,1999-Volkov-PhyRept,1990-Bizon-PRL,2014-Herdeiro-PRL,2014-Babichev-JHEP,2014-Sotiriou_Zhou-PRD,2014-Charmousis_Papantonopoulos-JHEP,2015-Herdeiro.Radu-IJMPD,2016-Barrientos.etal-EPJC}.} 
The field equations for a generic field-theoretic extension to GR take the following compact form~\cite{2010-DeFelice.Tsujikawa-LivingRev.Rel.,2010-Sotiriou.Faraoni-Rev.Mod.Phys.,2012-Clifton.etal-Phys.Rept.,2016-Joyce.etal-ARN,2017-Nojiri.etal-Phys.Rept.}: 
\begin{equation}
{\cal G}^{\rm (M)}_{\mu\nu} = 8 \pi G \, T_{\mu\nu} \, ,
\end{equation}
where ${\cal G}^{\rm (M)}_{\mu\nu}$ is the modified Einstein tensor, and $T_{\mu\nu}$ is the stress-tensor of the minimally coupled matter fields to gravity. Irrespective of the gravitational field equations, demanding the local conservation of the energy-momentum tensor $\nabla^{\mu} T_{\mu\nu}  = 0$, leads to the following generalized contracted Bianchi identity~\cite{1993-Hamity.Barraco-GRG,2016-Tian-GRG}:
\begin{equation}
\nabla^{\mu} {\cal G}^{\rm (M)}_{\mu\nu} = 0 \, .  
\label{eq:Bianchiidentity}
\end{equation}
In GR, the above Bianchi identify vanishes identically because the Einstein tensor has zero divergence ($\nabla^{\mu} {G}_{\mu\nu} = 0$). However, this is not the case for modified gravity theories. To see this, let us consider the following quadratic gravity action~\cite{1978-Stelle-GRG,2021-Nenmeli.etal-PLB}:
\begin{equation}
S= \frac{1}{16 \pi G}  \int d^{4} x \sqrt{-g}\left(R - \alpha R^{2} + \beta R_{\mu \nu} R^{\mu \nu}  \right) \, , 
\end{equation}
where $\alpha$ and $\beta$ are coupling constants. The generalized contracted Bianchi identity for the above action becomes:
\begin{align}
- \nabla^{\mu}G_{\mu\nu} & =   {\beta \, \nabla^{\mu} \left(-\frac{1}{2} R_{\rho \sigma} R^{\rho \sigma} g_{\mu \nu}-\nabla_{\nu} \nabla_{\mu} R-2 R_{\rho \nu \mu \sigma} R^{\sigma \rho}+\frac{1}{2} g_{\mu \nu} \square R+\square R_{\mu \nu}\right)} \nonumber \\ 
& +  \alpha \, \nabla^{\mu} \left(\frac{1}{2} R^{2} g_{\mu \nu}-2 R R_{\mu \nu}-2 \nabla_{\nu} \nabla_{\mu} R+2 g_{\mu \nu} \square R\right) 
\end{align}
As mentioned above, since the Einstein tensor has zero divergence, the LHS of the expression vanishes. By definition, the Kerr solution satisfies the condition $R_{\mu\nu} = 0$ and (hence, $R = 0$). In other words, any modified gravity model that satisfies $R_{\mu \nu} =0$ will have the Kerr metric as a black hole solution. The above expression shows that the Kerr solution satisfies the quadratic gravity field equation and the generalized Bianchi identity expression.

Note that the discussion in Sec.~\eqref{sec:KerrSchild}  is not restricted to any form of $f(R)$. However, in Sec.~\eqref{sec:Solutions}, as we obtain the rotating black hole solutions for the modified Einstein's equations, a specific $f(R)$ form becomes necessary. Consequently, the solutions presented correspond to this analytical $f(R)$ form. Nevertheless, as demonstrated in Section~\eqref{sec:KerrSchild}, the results should be valid for any $f(R)$.

However, no-hair theorem in GR states that black-holes are \emph{uniquely} described by the Kerr metric. Thus, in modified theories of gravity like $f(R)$ and quadratic gravity, the question arises: Is Kerr a unique solution to the modified gravity theories? From the above expression, we see that the  $R_{\mu\nu} = 0$ is a trivial solution to the above equation and non-trivial solutions can exist for such theories. 
In this work, we explicitly showed this for a class of $f(R)$. {Thus our analysis provides sufficient evidence 
that merits extending the no-hair theorem by including the coupling constants
~\cite{2012-Garcia.etal-PRD,2012-Chrusciel.etal-LRR}.} This is currently under investigation.

In GR, the no-hair theorem demands that the only memory of the structure and composition of any object that collapses to form a stationary BH is embodied in the mass ($M$) and angular momentum ($a$), with any residual hair rapidly radiated away during the collapse process. In the case of $f(R)$, our analysis shows that this is affected by the choice of $N_2$. Since $f(R)$ contains fourth-order derivatives, we need four boundary (initial) conditions to fix these constants. Here, we only demanded the asymptotic properties of the line element; hence, two of the constants $(C_4, N_2)$ are still undetermined. One possibility is to impose boundary conditions on the horizon or ergosphere. This is currently under investigation. 

The current understanding is that not all quantum field theories need to be renormalizable in order to be predictive and useful. This understanding gave rise to the idea of effective field theories. The effective field theory is a simplified model that accurately describes physical phenomena at a certain energy scale, while ignoring higher energy effects. In that context, non-renormalizable theories can still have predictive and practical value if they are seen as effective field theories. This is especially true when examining processes that occur at energy scales below a specific cutoff scale, represented by $M$.

The cutoff scale $M$ represents the energy scale beyond which the theory is no longer valid, and it serves as a boundary for the domain of applicability of the theory. At energy scales much lower than the cutoff scale ($E \ll M$), only a finite number of Feynman diagrams contribute to physical observables, making calculations tractable and predictions reliable within the precision of measurements. In that context the $f(R)$ model we have considered is an effective theory whose validity is set by the ratio $\alpha_1/\alpha_0$. In the limit $\alpha_1/\alpha_0 \ll 1$, $f(R)$ reduces to general relativity. It is an improved model compared to GR for $\alpha_1/\alpha_0 \sim 1$. However, our analysis shows that this is not sufficient to remove the singularity at the origin.

In Ref.~\cite{2018-Canate-CQG}, the author showed that non-trivial BHs do not exist in four-dimensional asymptotically flat, static, and spherically symmetric or stationary axisymmetric by imposing two conditions on $f(R)$. Although our $f(R)$ model \eqref{f(R)2ndorder} satisfies both these conditions, we have shown that multiple SRBH solutions exist. One possible reason for this discrepancy is that the generalized Bianchi identity \eqref{GeneralizedBianchi} leads to two conditions $\partial_R^2[f(R)] =0$ or $R_{\mu\nu} \nabla^{\mu} R = 0$. Trivial solutions like Ricci-flat space-times will automatically satisfy the above condition. However, the generalized Bianchi identity also implies that non-trivial solutions can exist. Here we have shown that the non-trivial solution with non-zero $\mu(\rho)$ decay to unity at asymptotic infinity.

We have analyzed some of the kinematical properties of the SRBHs and have shown that the horizon structure is different from that of Kerr. In addition, we have shown that the circular orbits for the SRBHs in $f(R)$ are smaller than that of Kerr. This means that the inner-most stable circular orbit for BHs in $f(R)$ is smaller; hence, the shadow radius might also be smaller. Thus, ngEHT can potentially be used to constrain $\kappa^2$. This is currently under investigation.

The next-generation GW detectors like the Einstein telescope and Cosmic explorer have higher sensitivity than the current LIGO-VIRGO-KAGRA detectors. These detectors will be highly sensitive in the quasi-normal mode (QNM) regime and can probe the QNM structure accurately in the strong gravity regime. It will be interesting to study the QNM frequencies for these non-trivial SRBH solutions and can provide strong constraints on the coupling constants~\cite{2016-Cardoso.Gualtieri-CQG}. This is currently under investigation.

We obtained the SRBH solutions by imposing the condition $W_2(\rho) = 0$. It will be interesting to know whether one can obtain non-singular BH solutions from the other branch
$W_1(\rho)= 0$. Even the non-existence of non-singular BH solutions in this $f(R)$ model might explain the cause of the persistence of singularities in modified gravity models. This is under investigation.

\section*{Acknowledgements}
The authors thank K. Chandra, S. M. Chandran, A. Chowdhury, J. P. Johnson, A. Kushwaha and T. Parvez for discussions and  comments on the earlier draft. The authors thank A. Tharaka Rama for pointing out a generalization of our earlier calculations. The work of SS is supported by the SERB-Core Research Grant. 

\appendix
\markboth{}{}


\section{Status of no hair conjecture in modified gravity theories}
\label{sec:refs}

As mentioned in the Introduction, it has been shown that the \emph{standard} no-hair theorem fails in some cases, like Einstein–Yang–Mills theory and nonlinear electrodynamics, among others~\cite{2015-Herdeiro.Radu-IJMPD,2016-Barrientos.etal-EPJC}.  
 
In Ref.~\cite{1995-Kanti-PRD}, the authors numerically demonstrated the existence of black-hole solutions with non-trivial dilaton hair for a 4-D Effective superstring Action in the presence of Gauss-Bonnet quadratic curvature terms. 
It has been shown that this black hole is stable~\cite{1997-Kanti-PRD}. The same authors also showed numerically the existence of black hole solutions with non-trivial dilaton and Yang-Mills hair for the particular case of SU(2) gauge fields~\cite{1996-Kanti-PLB}. In Ref.~\cite{1997-Torii-PRD}, black holes in an effective theory superstring model contain the dilaton field, gauge field, and the Gauss-Bonnet coupling terms. These authors obtained spherically symmetric, asymptotically flat black hole solutions in the $D$-dimensional low-energy effective heterotic string theory, which contains Gauss-Bonnet and the dilaton coupling terms~\cite{2008-Guo-ProgTheoPhys}.

Similarly, dilatonic black holes have been shown to arise in
 one-loop corrected 4-D effective heterotic string theory containing Gauss-Bonnet couplings~\cite{2009-Pani_Cardoso-PRD}. These are interesting objects as a prototype for alternative yet well-behaved gravity theories that evade the no-hair theorem of GR but were proved to be stable against radial perturbations. These authors extended the analysis and obtained non-trivial, slowly rotating black holes in alternative theories of gravity~\cite{2011-Pani_Cardoso.et.al-PRD}. In Ref.~\cite{2011-Kleihaus_Eugen-PRL}, the authors numerically obtained generalizations of the Kerr black holes by including higher curvature corrections in the form of the Gauss-Bonnet density coupled to the dilaton. The same authors numerically obtained solutions for Einstein–Gauss-Bonnet–dilaton (EGBd) theory using a non-perturbative approach by directly solving the field equations~\cite{2016-Kleihaus_Eugen.et.al-PRD}. These stationary axially symmetric black holes are asymptotically flat.

Recently, in Ref.~\cite{2018-Antoniou_Kanti-PRL}, the authors numerically obtained black-hole solutions with a regular horizon in a general Einstein-scalar-GB theory with a coupling function $f(\phi)$. Thus, the authors showed that the existing no-hair theorems are easily evaded, and a large number of regular, black-hole solutions with scalar hair are then presented for a plethora of coupling functions $f(\phi)$~\cite{2018-Antoniou_Kanti-PRD}. For a class of extended scalar-tensor-Gauss-Bonnet (ESTGB) theories and mentioned that in this class of ESTGB theories, the authors showed the existence of new black hole solutions that are formed by spontaneous scalarization of the Schwarzschild black holes in the extreme curvature regime~\cite{2018-Doneva_Yazadjiev-PRL}. In Refs.~\cite{2018-Silva_Sotiriou_Berti-PRL,2018-Brihaye_Urrestilla-JHEP}, the authors identified a class of scalar-tensor theories with coupling between the scalar and the Gauss–Bonnet invariant that exhibits spontaneous scalarization for both black holes and compact stars. It has been shown that non-trivial black hole solutions exist in Einstein-scalar-Gauss-Bonnet theory in the presence of a cosmological constant $\Lambda$, either positive or negative~\cite{2018-Bakopoulos_Kanti-PRD}. See also, Refs~\cite{2019-Minamitsuji_Taishi-PRD,2019-Macedo_Sakstein_Berti_Sotiriou-PRD,2019-Doneva_Staykov_Yazadjiev-PRD,2019-Zou_Myung-PRD,2019-Cunha_Eugen-PRL}.

In Ref.~\cite{2018-Sultana_Demosthenes-GenReGra}, the authors showed that the no-hair theorem for spherically symmetric black holes does not exist for $R^2$ gravity. Our analysis in this work suggests that the no-hair theorem for modified gravity theories merits extending to include the coupling constants. 

\section{Slowly rotating black hole solution: Deformation linear in spin}
\label{App:A}

Consider the following SRBH line element:
\begin{align}\label{SRBHlinelement}
    ds^2 = -e^{\delta(\rho)} \,A(\rho)  d\tau^2 + \frac{d\rho^2}{A(\rho)} - 2 \, \chi \, B(\rho)\, \sin ^2(\theta)\,d\tau\,d\phi + \rho^2 d\Omega^2
\end{align}
where $A(\rho)$ and $\delta(\rho)$ are unknown functions of the reparameterized radial coordinate $\rho$. We use two different  --- modified Bianchi identity \eqref{GeneralizedBianchi} and modified equations of motion~\eqref{ModifiedFEq} --- approaches to obtain the solution.

\subsection{Approach I: Bianchi Identity}

As mentioned earlier, the generalized Bianchi identity \eqref{GeneralizedBianchi} leads to two conditions. First, $f''(R) = 0$ leads to a trivial solution and corresponds to GR. The second condition can potentially lead to non-trivial solutions:
\begin{align}
     R_{\mu\nu} \nabla^{\mu}R = 0 \quad \mbox{or} \quad 
     R^{\beta}_{\gamma} \nabla_{a}R = 0 
\end{align}
For the SRBH line element \eqref{SRBHlinelement}, using the fact that Ricci scalar is time-independent, i. e. $\partial_{t} R = 0$ and Ricci tensor components are non-zero, the above four equations can be combined to give the following expression: 
\begin{align}\label{f(R)Condition-Rot}
\left[R_{\rho \rho} + R_{\rho \theta}\right]\,g^{\rho \rho}\partial_{\rho}R +  \left[R_{\theta \rho} + R_{\theta \theta}\right]\,g^{\rho \rho}\partial_{\rho}R = 0 \, .
\end{align}
Substituting the line-element \eqref{SRBHlinelement} in the above expressing and ignoring all terms containing $\chi^q$ for $q>1$, we can write the above expression as a product of two second-order differentials in $A(r)$ and $\delta(r)$:
\begin{equation}
\label{eq:Linearorder-Bianchi}
2 A^2(\rho) \rho^4 \, \times \, 
{\cal E}_1[\delta(\rho), A(\rho)] \times {\cal E}_2[\delta(\rho), A(\rho)] = 0
\end{equation}

where,
 {\small
 \begin{align}
{\cal E}_1[\delta(\rho), A(\rho)] & = 
\delta''(\rho) + \frac{\delta'(\rho)^{2}}{2} + 
\left[
\frac{3}{2} \,\delta'(\rho) +  1
+ \frac{2}{\rho} \right] \frac{A'(\rho)}{A(\rho)}  \\
{\cal E}_2[\delta(\rho), A(\rho)] & = 
\delta'''(\rho) 
+ \left[\delta'(\rho) + \frac{5}{2} \frac{A'(\rho)}{A(\rho)} + \frac{2}{\rho} \right]\delta''(\rho) 
+ \frac{1}{2} \frac{A'(\rho)}{A(\rho)} \delta'(\rho)^{2} 
+ 2 \left[ \frac{1}{\rho} \frac{A'(\rho)}{A(\rho)}
- \frac{1}{\rho^2} \right]\, \delta'(\rho)  \,
\nonumber\\&
~~~~+  \frac{A'''(\rho)}{A(\rho)} 
+ \left[\frac{3}{2} \delta'(\rho) + \frac{4}{\rho} \right] \frac{A''(\rho)}{A(\rho)}
- \frac{2}{\rho^2} \frac{A'(\rho)}{A(\rho)} 
- \frac{4}{\rho^3} \left[1 - \frac{1}{A(\rho)} \right] \, ,
\end{align}
}
and prime denotes derivative w.r.t $\rho$. This is an interesting result regarding which we want to discuss the following: First, the above condition \eqref{eq:Linearorder-Bianchi} is valid for any $f(R)$ model and any form of stress-tensor. This is because the Bianchi identity \eqref{GeneralizedBianchi} is derived by setting $\nabla_{\mu} T^{\mu\nu} = 0$. Second, the above condition is independent of 
$B(\rho)$. This implies that at linear-order in spin $B(\rho)$ is arbitrary. In Ref.~\cite{2020-Xavier.etal-CQG}, for spherically symmetric space-times, the authors showed that $\delta(\rho)$ could take infinitely many values. Here we have shown that $B(\rho)$ also can take infinite values. To know whether this is indeed the case, we proceed to obtain the condition on $B(\rho)$ 
using the equations of motion~\eqref{ModifiedFEq}.

\subsection{Approach II: Equations of Motion}

To confirm the same, we use the equations of motion~\eqref{ModifiedFEq} 
for model given in Eq.~\eqref{f(R)2ndorder} for $p = 2$. Substituting the line element \eqref{SRBHlinelement},
in Eq.~\eqref{SecondOrderModified-General}, up to leading order in $\chi$, we obtain the following form of the modified field equations: 
\begin{subequations}
\label{eq:Linord-ModEqs}
\begin{align}
    &{\cal G}^{\tau}_{\tau} \equiv {\cal T}_{1}[A(\rho),\delta(\rho)] = 0\\
    &{\cal G}^{\tau}_{\phi} \equiv {\cal T}_{2}[A(\rho),\delta(\rho),B(\rho)] = 0\\
    &{\cal G}^{\rho}_{\rho} \equiv {\cal T}_{3}[A(\rho),\delta(\rho)] = 0\\
    &{\cal G}^{\theta}_{\theta} \equiv {\cal G}^{\phi}_{\phi} \equiv {\cal T}_{4}[A(\rho),\delta(\rho)] = 0\\
    &{\cal G}^{\phi}_{\tau} \equiv {\cal T}_{5
    }[A(\rho),\delta(\rho),B(\rho)] = 0
\end{align}
\end{subequations}
where ${\cal T}_{1}, {\cal T}_{3}, {\cal T}_{4}$ and ${\cal T}_{5}$ are functions of $A(\rho)$ and $\delta(\rho)$ while ${\cal T}_2$ is a function of $A(\rho), B(\rho)$ and $\delta(\rho)$. Specifically, like in spherically symmetric case~\cite{2020-Xavier.etal-CQG}, ${\cal T}_{1}$ and ${\cal T}_{4}$ have up to $4^{th}$ order derivatives of $A(\rho)$ and $\delta(\rho)$, ${\cal T}_{3}$ have up to $3^{rd}$ order derivatives of $A(\rho)$ and $\delta(\rho)$, ${\cal T}_{2}$ and ${\cal T}_{5}$ have up to $3^{rd}$ order terms of both $A(\rho)$ and $\delta(\rho)$ and up to $2^{nd}$ order terms for $B(\rho)$. It is possible to rewrite ${\cal T}_{5}$ in terms of the other components.

Since the equations of motion contain up to fourth-order derivatives of $A(\rho)$ and $\delta(\rho)$, an exact solution will contain up to four independent constants. Using the procedure used in Ref.~\cite{2020-Xavier.etal-CQG} for the spherically symmetric space-times, by combining ${\cal T}_1, {\cal T}_3$ and ${\cal T}_4$, we obtain the following expression: 
\begin{align}
    \frac{3\,w_{1}(\rho)\,w_{2}(\rho)}{4\,\rho^{3}\,A^2(\rho)\,\,\,[\Phi(\rho) -2]\,[\Phi(\rho) + 4]} =0, 
\label{EOMfinalEq}
\end{align}
where, $\Phi(\rho) = \rho(\delta'(\rho) + [\ln A(\rho)]')$, 
{\small
\begin{align}
 w_{1} =& A(\rho)\left[2\,[\ln A(\rho)]'(2-\rho) +\frac{\Phi(\rho)}{2}\left(\rho\,[\ln A(\rho)]' + 2\right) + \rho\,\Phi'(\rho) + 2\right] - 2\left[1 + \frac{\alpha_{0}\rho^2}{2\alpha_{1}}\right] \\
    w_{2} =& A^2(\rho)\,\rho^2 \left(\frac{\Phi(\rho)}{\rho} - [\ln A(\rho)]'\right)^2\left(\frac{3}{2}\rho [\ln A(\rho)]' +\Phi(\rho) + 1\right) - 2A(\rho)\left(A(\rho) -1\right)\left([\ln A(\rho)]' -2\right) \nonumber \\&
    +\,62 A(\rho)\rho^2\left(\left[\frac{\Phi(\rho)}{\rho}\right]' + ([\ln A(\rho)]')^2\right)\left(\Phi(\rho) + \frac{4}{3}\right)-\frac{2\alpha_{0}\,\rho}{3\,\alpha_{1}}\left(2\rho^2\,A'(\rho) - A(\rho)\Phi(\rho) + \rho\right)
    \nonumber \\& 
     + \left(\frac{\Phi(\rho)}{\rho} - \frac{\rho[\ln A(\rho)]'}{2}\right)\left(3\rho\left(\rho [A'(\rho)]^2 + 2[\ln A(\rho)]'\right)+4\,A(\rho)(A(\rho)+1)\right)
\end{align}
}

Demanding a non-trivial solution to be satisfied for any finite value of $\rho$ leads to the following conditions 
\begin{align}
w_{1} = 0~\mbox{or}~w_{2} = 0~{\rm while}~~\Phi(\rho) \neq 2,\, -4 
\label{ConditionPhi}
\end{align} 
Thus, we have rewritten the fourth-order differential equation as a product of two second-order non-linear differential equations. This allows for obtaining an exact slowly-rotating black hole solution. It is important to note that the above differential equations impose conditions on $A(\rho)$ and $\delta(\rho)$ only and does impose any condition on $B(\rho)$. To obtain a condition on $B(\rho)$, we eliminate the third-derivative in $A(\rho)$ or third-derivative using ${\cal T}_2$ and ${\cal T}_5$. This leads to the following equation:
\begin{align}
\frac{w_{1}(\rho)\,w_{3}(\rho)}{2 [\rho\,B'(\rho) - 2\,B(\rho)]\,\left[B(\rho)\,\Phi(\rho) - \rho\,B'(\rho)\right]} = 0 \label{EOMcondiwithp(r)}
\end{align}
where 
{\small
\begin{align}
w_{3} =&\, 
\left[\frac{B'(\rho)}{B(\rho)} 
- \frac{2}{\rho} \right] \left(\left[\frac{\Phi(\rho)}{\rho}\right]' + ([\ln A(\rho)]')^2 + \left[\frac{\Phi(\rho)}{\rho} - [\ln A(\rho)]'\right]\left[\frac{\Phi(\rho)}{\rho} - [\ln 
 A(\rho)]'  + 
 \frac{2}{A(\rho)} \right] + 2\right) \nonumber \\
& - \frac{B''(\rho)}{\rho B(\rho)} 
\left(\Phi(\rho) - 2\right) + 
\frac{2}{\rho^2} B(\rho)\left(
\Phi(\rho) - \rho [\ln A(\rho)]'\right) + \frac{2}{\rho^2} \left([\ln A(\rho)]' - \frac{1}{B(\rho)} \right)
\end{align}
}
Here again, demanding a non-trivial solution to be satisfied for any finite value of $\rho$ leads to the following conditions 
\begin{align}
w_{1} = 0~\mbox{or}~w_{3} = 0~~{\rm while}~~B(\rho) \neq 0,\,\rho^2,\, \exp\left[\int \Phi(\rho) d(\ln\rho) \right] \, .
\label{ConditionPhi2}
\end{align}
From Eqs. (\ref{ConditionPhi}, \ref{ConditionPhi2}), we infer that $\omega_1 = 0$ will satisfy all the modified Einstein's equations \eqref{eq:Linord-ModEqs}.  Rewriting $\Phi(\rho)$ as:
\begin{align}
\delta'(\rho) + [\ln A(\rho)]' = \nu(\rho)\,\,\text{and}\,\,\nu(\rho)\neq\frac{-4}{\rho}\,\text{or}\,\frac{2}{\rho} \, ,
\end{align}
we see that given a $\nu(\rho)$, we 
have a relation between $A(\rho)$ and $\delta(\rho)$. More specifically, for a given $\Phi(\rho)$, we can obtain $A(\rho)$ by solving $w_1 = 0$. However, there are infinite choices for $\nu(\rho)$. For instance, solving $w_1 = 0$, leads to: 
\begin{align}
\label{eq:A-Linearchi}
   A(\rho) = 1 - \frac{C_1}{\rho} + \frac{C_{0}}{\rho^2} + \frac{\rho^2}{12 \kappa^2}
\end{align}
Hence, we have infinite vacuum solutions with infinite forms of $\delta(\rho)$. Moreover, since obtaining $A(\rho)$ does not constrain the form of $B(\rho)$, we have an infinite number of SRBH solutions~\eqref{SRBHlinelement}. 

\section{Slowly rotating black hole solution: Quadratic order in spin}
\label{App:B}

This appendix details calculations of the SRBH solution at the quadratic order in the spin parameter $(\chi)$. We do this in two steps. First, we set $\mu(\rho) = 0$ and obtain the analytical expressions for $U(\rho)$ and $V(\rho)$. We then use the form of $U(\rho)$ in 
Eq.~\eqref{eq:Secord-ModEqs} to obtain the form of $\mu(\rho)$.

\subsection[Testof]{Obtaining $U(\rho)$ by setting $\mu(\rho) = 0$ in Eq.~(\ref{EQM2ndorderForm})}
 
Since Eq.~\eqref{EQM2ndorderForm} is highly non-linear and depends on three unknown functions, we first consider the standard form of Eq.~\eqref{Slow_Kerr} by setting $\mu(\rho) = 0$. In this case, we have:
\begin{align}
W_{1}(\rho)&= \left[3\,\rho^2 U''(\rho)  
-6 (U(\rho) - 1) \right] \left[\rho U'(\rho) + 2 U(\rho)\right] 
+  \frac{\alpha_{0}}{\alpha_{1}} \rho^2 
 \left[\rho U'(\rho) - 2 U(\rho)\right] \\
W_{2}(\rho) &= \rho^2\, U''(\rho) + 4\,\rho\,U'(\rho) + 2\,U(\rho) -\rho^2 \frac{\alpha_0}{\alpha_1} - 2 
\end{align}
Demanding that Eq.~\eqref{EQM2ndorderForm} is satisfied at each order of $\chi$~\cite{1967-Hartle-ApJ,1968-Hartle.Thorne-ApJ}, we have the following conditions:
\begin{equation}
W_1(\rho) = 0~~{\rm or}~~W_2(\rho) = 0; ~W_0(\rho) = 0   
\end{equation}
Setting $W_2(\rho) = 0$, we obtain the following solution for $U(\rho)$:
\begin{align}\label{SolofU}
U(\rho) =  1 - \frac{C_1}{\rho} + \frac{C_{0}}{\rho^2} + \frac{\rho^2}{12 \kappa^2} 
\end{align}
where $C_0, C_1$ and $C_2$ are constants. Note that this is identical to Eq.~\eqref{eq:A-Linearchi} we derived in the case of deformation linear in spin. For further analysis we consider $C_2 = 0$. Having determined $U(\rho)$, our next step is to determine $V(\rho)$. For this, we can use any of the following assumption that all the components of the modified Einstein tensor ($ {\cal G}^{\mu}_{\nu}$) (\ref{eq:Secord-ModEqs} )should vanish asymptotically, which will give;
\begin{equation}
V(\rho) = \sum_{j=0}^{\infty}\frac{C_{j+2}}{\rho^{j+2}} \, .
\label{SolofV}
\end{equation}
Where $C_{3+j}$ are constants. In order to fix these constants, we substitute the above form of $U(\rho), V(\rho)$ and $\mu(\rho)$ in Eq.~\eqref{EQM2ndorderForm} and impose the condition that at asymptotic infinity ($\rho \longrightarrow \infty$), the modified Einstein tensor ($ {\cal G}^{\mu}_{\nu}$) components should vanish. They are of the following form:
\[
T_i \longrightarrow \chi^2\, T^{(II)}_{i}\left(\frac{1}{\rho^2}\right);~~
T_4 \longrightarrow \chi^2\, T^{(II)}_{4}\left(\frac{1}{\rho}\right)\,,
\]

\subsection[test]{Obtaining $\mu(\rho)$}

Having obtained $U(\rho)$ and $V(\rho)$ from \eqref{SolofU} and \eqref{SolofV} , we now determine the form of $\mu(\rho)$ by solving $W_{2}(\rho) = 0$. Substituting the above form of $U(\rho)$ (and not fixing the value of $C_1$), we have:
\begin{align}
\mu''(\rho) + \frac{\mu'(\rho)^{2}}{2} + \frac{\left[5\,\rho^3 + 24\,\kappa^2\,\rho - 12 \, C_1 \, \kappa^2\right]}{\rho\,\left[\rho^3 + 12\,\kappa^2\,\rho - 12\,C_{1}\,\kappa^2\right]} \mu'(\rho) = 0
\end{align}

Solving for $\mu(\rho)$ we get a solution in the integral form:
\begin{align}
\label{eq:musol}
\frac{\mu(\rho)}{2} = 
\ln \left[{N_1} \left(\int \rho^{-1/2}\left(\rho^3 + 12\,\rho\,\kappa^2 - 12\,C_{1}\,\kappa^2\right)^{-3/2} \,d\rho\right) + {N_2} \right]
\end{align}
where $N_1,N_2$ are constants of integration. In the asymptotic limit, the above integral form reduces to: 
\begin{align}
\frac{\mu(\rho)}{2} \propto 
\ln \left[1 - \frac{5 \, N_1}{\rho^4}\right] \, ,
\end{align}
where we have scaled the constant and positive definiteness of $\exp[\mu(\rho)]$ demands that $N_1 < 0$. Substituting the above integral form of  $\mu(\rho)$ \eqref{eq:musol}, and  $U(\rho), V(\rho)$ from from \eqref{SolofU} and \eqref{SolofV} all the modified Einstein tensor components vanish as
\begin{align}
T_i \propto \chi^2\, T^{(II)}_{i}\left(\frac{1}{\rho^2}\right);~~
T_4 \propto \chi^2\, T^{(II)}_{4}\left(\frac{1}{\rho}\right)\,,
\end{align}
where $i = 1,2,3$. 

\section[test]{Comparing the two forms of $V(\rho)$} \label{App:C}

\begin{figure*}[!htb]
\begin{minipage}[b]{.5\textwidth}
	\centering
    \subcaptionbox{  \label{fig:LzEchi+ve-2Sol} }{
\includegraphics[width=\columnwidth]{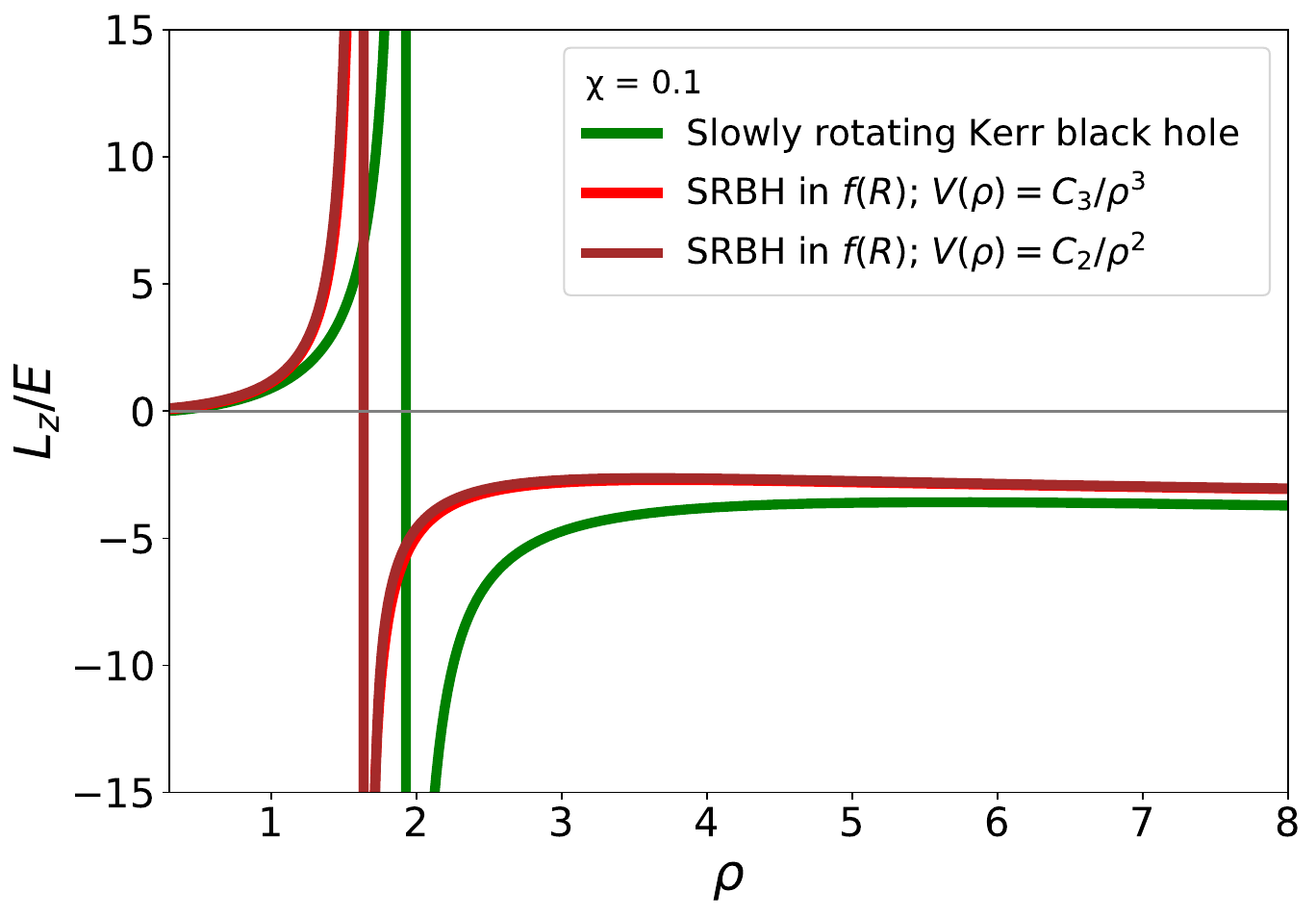}   }

\end{minipage}\hfill
\begin{minipage}[b]{.5\textwidth}
	\centering
    \subcaptionbox{  \label{fig:LzEchi-ve-2Sol} }{
\includegraphics[width=\columnwidth]{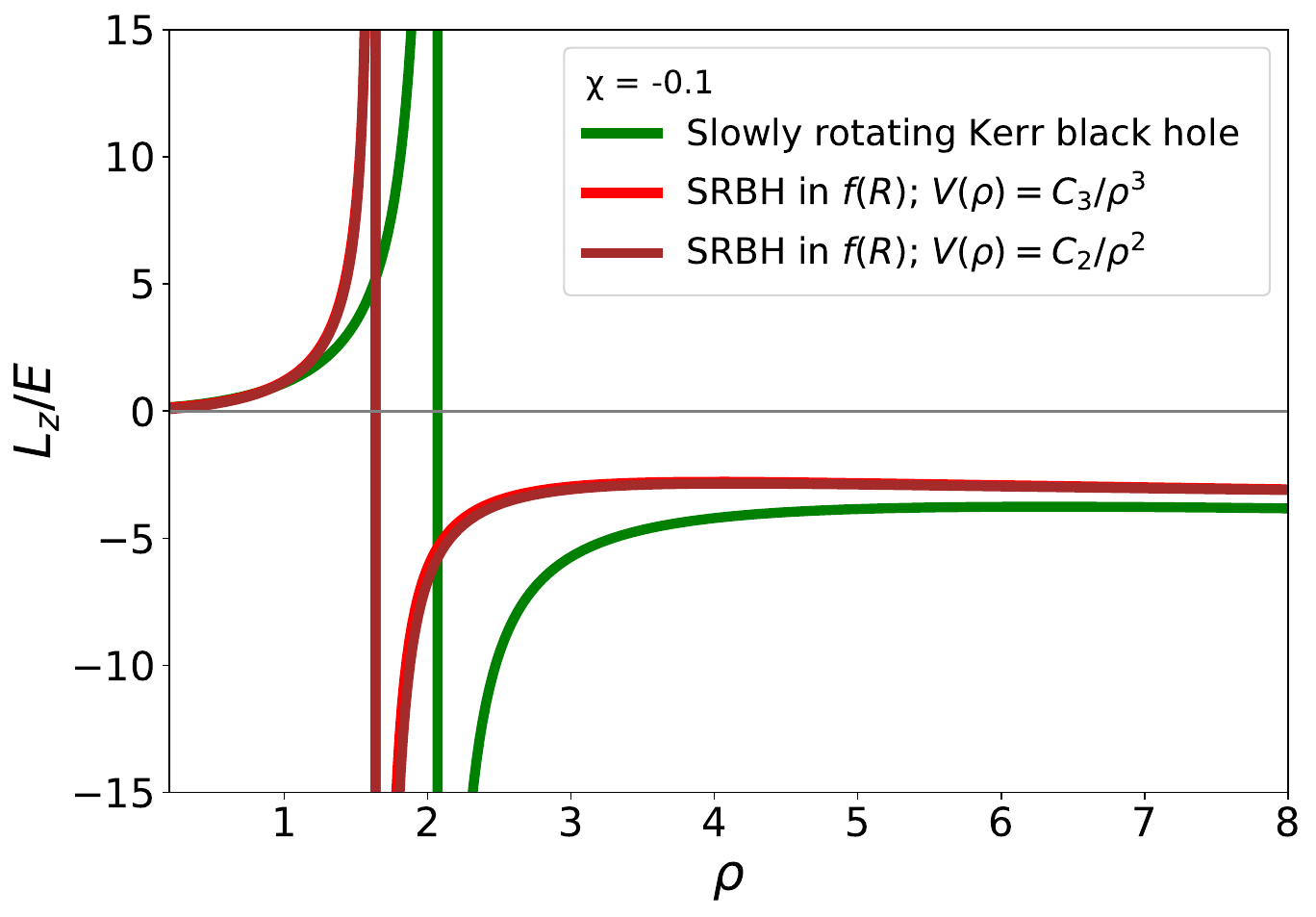}  }

\end{minipage} 
\caption{Plot of $L_z/E$ of a test particle in equatorial circular orbits as a function of $\rho$ for $\kappa^2 =1$ and $C_1 =2$. We also plot for SR Kerr solutions. (a) For  $\chi = 0.1$. (b) For  $\chi = - 0.1$.}
\label{fig:LzE-2Sol}
\end{figure*}

\begin{figure*}[!htb]
\begin{minipage}[b]{.5\textwidth}
	\centering
    \subcaptionbox{ \label{fig:EffV-2Sol} }{
\includegraphics[width=\columnwidth]{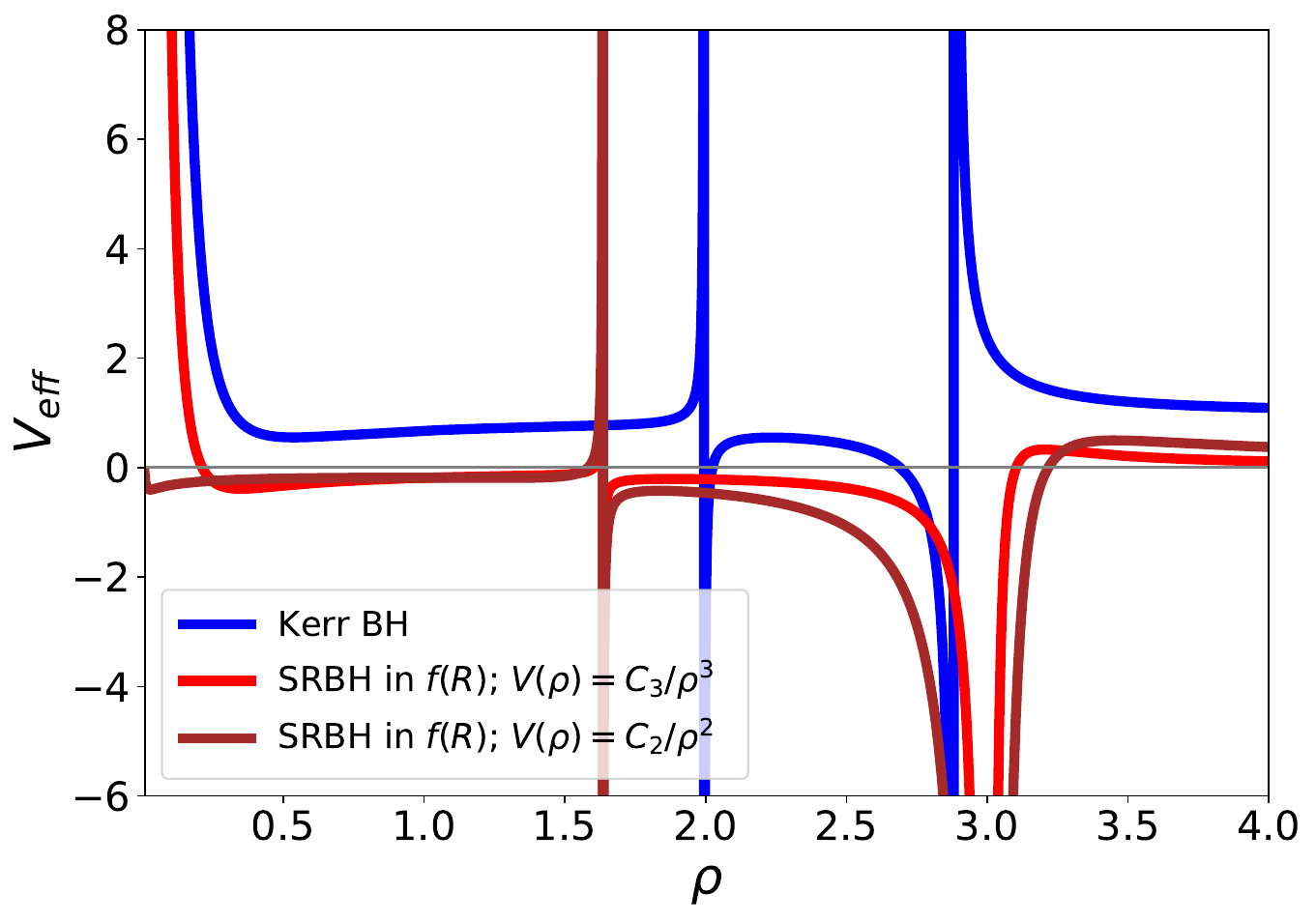}  }

\end{minipage}\hfill
\begin{minipage}[b]{.5\textwidth}
	\centering
    \subcaptionbox{ \label{fig:zamo-2Sol} }{
\includegraphics[width=\columnwidth]{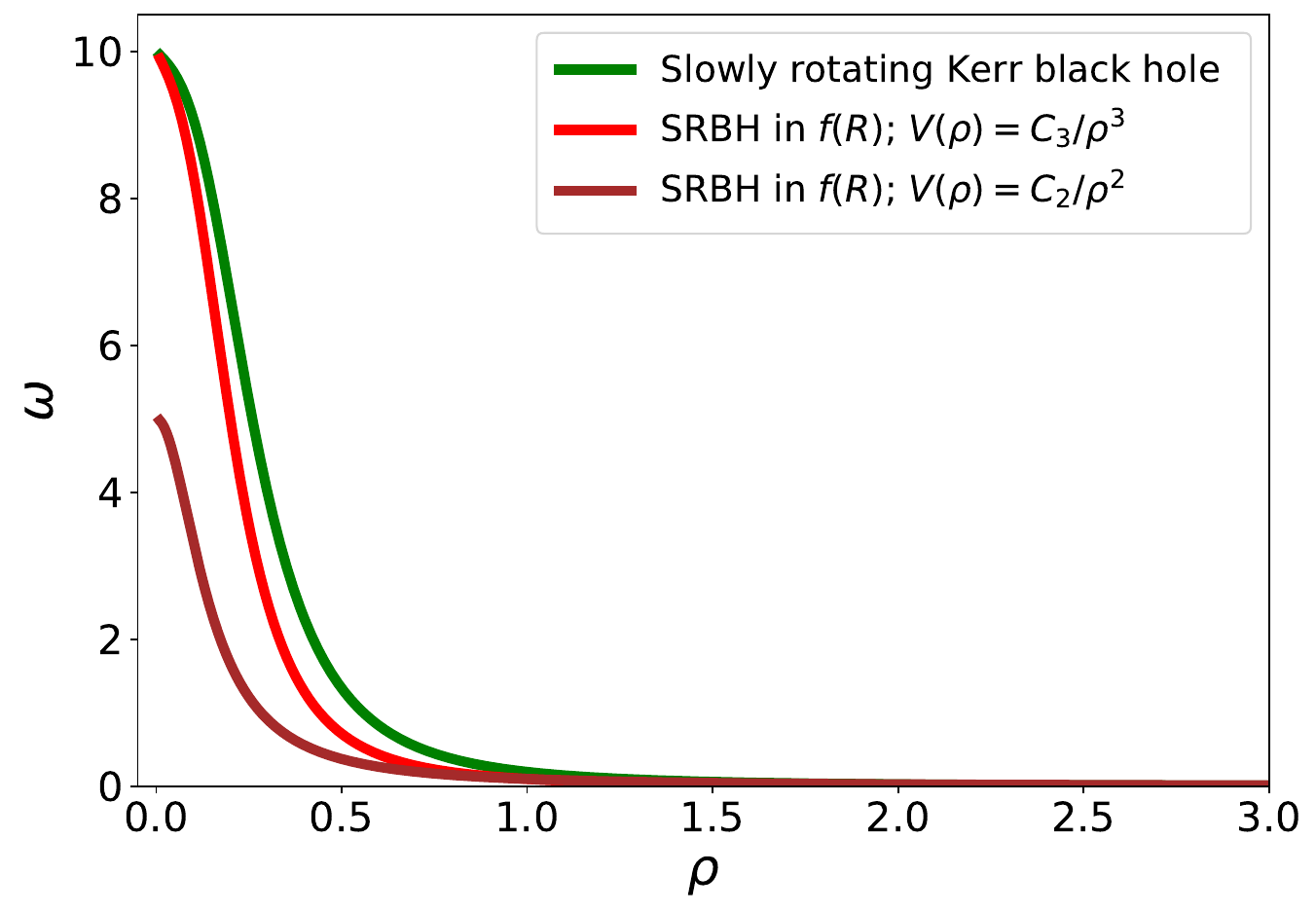}  }

\end{minipage} 
\caption{(a) Plot of $V_{\rm eff}$ of a test particle in the equatorial plane as a function of $\rho$. We set $C_1 = 2$.
(b) Plot of the equatorial plane ZAMO's angular velocity as a function of $\rho$. For line-element  \eqref{Slow_Kerr}, we have set $\kappa^2 = 1, \chi = 0.1$ and $C_2 = C_3=1$. We also plot for SR Kerr and
exact Kerr.}
\label{fig:VeffZamo-2Sol}
\end{figure*}

In this appendix, we compare the kinematical properties of the SRBH for two different forms of $V(\rho)$ obtained in Eq.~\eqref{eq:Solof_V}. Specifically, we consider the following two forms: 
%
\begin{align*}
    V(\rho) = \frac{C_2}{\rho^2} \quad \text{and} \quad V(\rho) = \frac{C_3}{\rho^3}
\end{align*}
As mentioned in Sec.~\eqref{sec:Properties}, both of forms of $V(\rho)$ correspond to a SRBH with the same event horizon Eq.~\eqref{eq:horizon}. 
Fig.~\eqref{fig:LzE-2Sol} compares $L_z/E$ for the two cases for different values of $\chi$. Fig.~\eqref{fig:EffV-2Sol} compares the effective potential $V_{eff}$ and ZAMO for the two cases for $\chi = 0.1$. We see that the features of all these kinematical quantities are the same for the two cases. Hence, it is not possible to distinguish the two solutions.
%

%
%

\section{Comparing our solution with asymptotically non-flat rotating black hole}\label{App:D}

In literature,  the asymptotically non-flat rotating black hole in GR is given by~\cite{2009-Lu_Pope-IOP,1973-Demianski-Acta_Astro,2005-Gibbons-JGP}:
\begin{align}\label{eqnkrasinski}
ds^{2}&= -\left( 1- {\frac {2\,M\,r}{r^2+a^2 \cos^{2} \theta }}+\frac{1}{3}\,\Lambda'\,\left(r^2+a^2 \cos^{2} \theta\right) \right) dt^2 +\left[ {\frac {r^2+a^2 \cos^{2} \theta}{\frac{\Lambda'\,r^2}{3}\,\left(a^{2}+r^{2} \right)+r^2-2\,Mr+a^{2} } } \right] dr^2
\nonumber\\&
- 2\,a\,\sin^{2} \theta \,\left( {
\frac {2 M\,r}{r^2+a^2 \cos^{2} \theta}}-\frac{\Lambda'}{3}\left( r^2+a^2 \right) \right)\,dt d\phi + \left( {\frac {r^{2}+a^2 \cos^{2} \theta }{1-\frac{\Lambda'\,a^{2}\,\cos^{2}\theta}{3}
}} \right) d\theta^2
\nonumber\\&
+ \sin^{2} \theta \left( 
\,{\frac {2\,M\,r\,a^{2} \sin\theta^{2} }{r^2+a^2 \cos^2\theta}}+(r^2+a^2)\left(1-\frac{\Lambda'\, a^2}{3} \right) \right)d\phi^2
\end{align}
where $\Lambda'$ is the cosmological constant with dimension $[L]^{-2}$. We can re-write the above metric with dimensionless quantities: dimensionless time $\tau = t/M$, dimensionless radial coordinate $\rho = r/M$, the dimensionless spin parameter $\chi = a/M$ and dimensionless cosmological constant $\Lambda = M^2\,\Lambda' $. On applying this transformation, in the slow rotation approximation up to second order ($\chi$), we get:
\begin{align}\label{Asymp_Slow_Kerr}
    ds^2 = &- \left(1 - \frac{2}{\rho} + \frac{\Lambda \rho^2}{3} + \chi^2\,\cos^2(\theta)\left[\frac{6 +\rho^3\,\Lambda}{3\,\rho^3}\right]\right)\,d\tau^2  - \frac{2 \,\chi\,\sin^2(\theta) [6 - \Lambda\rho^3]}{3\rho}\,d\tau\,d\phi\nonumber \\
    & + \left(\frac{3\rho^2[\Lambda\rho^3 + 3\rho -6] - \chi^2\,3\left(\cos^2(\theta)[\Lambda\rho^3 + 3\rho -6] - \Lambda\rho^3 - 3\rho\right)}{[\Lambda\rho^3 + 3\rho -6]^2\,\rho}\right)\,d\rho^2 
    \nonumber \\
    & + \left(\rho^2 + \chi^2\,\frac{\cos^2(\theta)\,[\Lambda\,\rho^2 +3]}{3}\right)\,d\theta^2 + \rho^2\,\sin^2(\theta)\,\left(1 + \frac{\chi^2}{\rho^2} + \chi^2\left[\frac{2\,\sin^2(\theta)}{\rho^3} - \frac{\Lambda}{3}\right]\right)\,d\phi^2
\end{align}
having event horizon ($g^{\rho \rho} =0$) at:
\begin{align}
    \rho_{H} = \frac{H_{dS}^{2/3}\,\Lambda^{-1/3} - \Lambda^{-2/3}}{H_{dS}^{1/3}}\quad ; \qquad H_{dS} = \sqrt{9 + \frac{1}{\Lambda}} + 3
\end{align}
%
%
%
with the radius of the ergosphere ($g_{\tau \tau} = 0$) of the form:
\begin{align}
    \frac{\Lambda\,\rho^5}{3} + \rho^3\left(1 + \frac{\chi^2\,\Lambda\,\cos^2(\theta)}{3}\right) - 2\rho^2 + 2\,\chi^2\,\cos^2(\theta) = 0.
\end{align}
It should be noted that the analytical form of our solution Eq.~(\ref{Slow_Kerr}) is similar to the slowly rotating form of the solutions in the literature. For more understanding, we now compare the metric Eq.~(\ref{Asymp_Slow_Kerr}) with Eq.~(\ref{Slow_Kerr}), where
%

\begin{equation}
U(\rho) = 1- \frac{C_1}{\rho} + 
\frac{\rho^2}{12 \, \kappa^2}; \qquad
V(\rho) = \frac{C_{3}}{\rho^3}; \qquad \mu(\rho) = 0 \, . 
\label{eq:SolofU-V2}    
\end{equation}
\begin{figure*}[!htb]
\begin{center}
\includegraphics[width=0.5\textwidth]{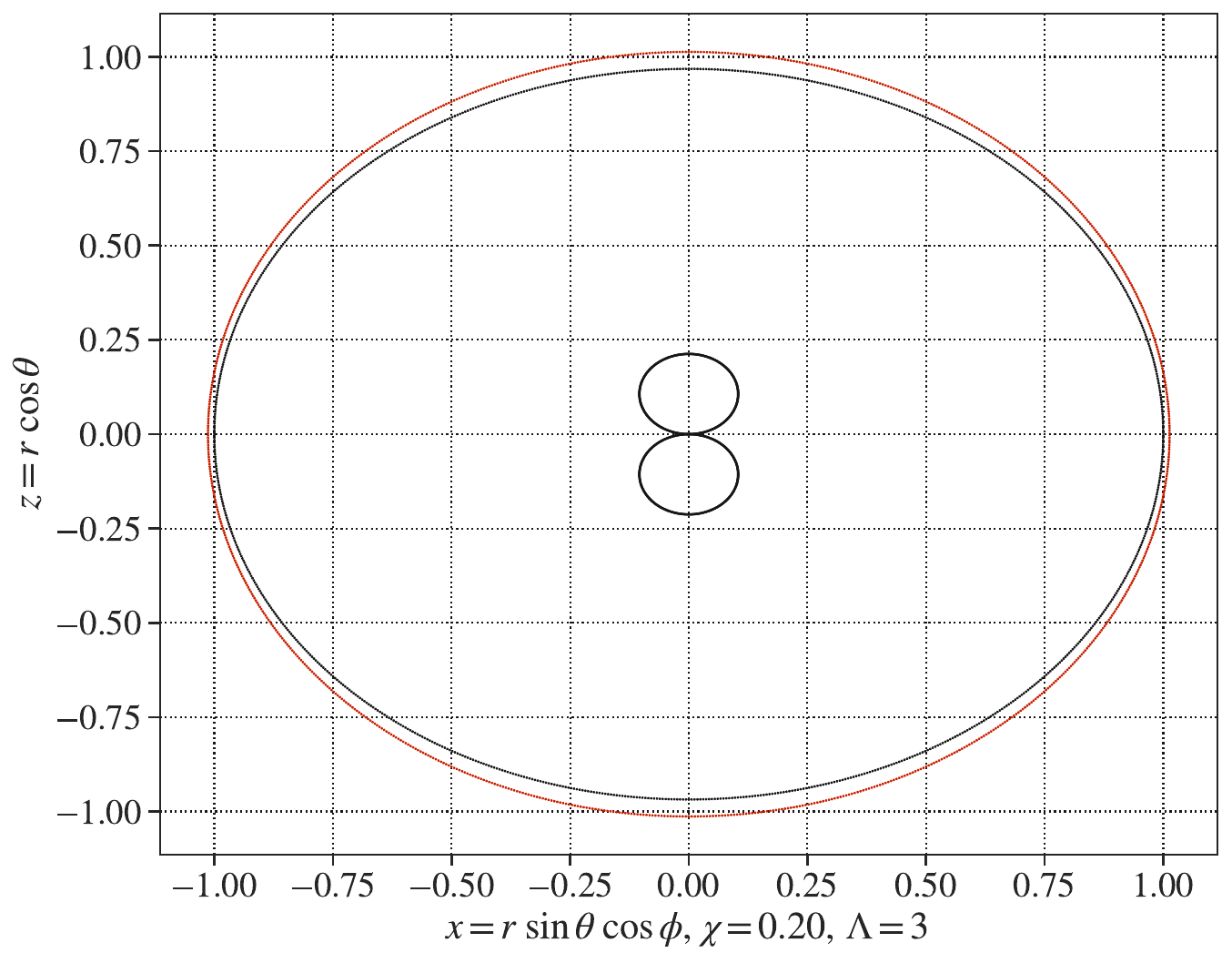}
\caption{Plot of ergosphere of Asymptotic de Sitter Kerr SRBH (in GR) for $\chi = 0.2$ and $\Lambda = 3$}
\label{fig:ergosphereNonflat}
\end{center}
\end{figure*}

\begin{figure*}[!htb]
\begin{minipage}[b]{.5\textwidth}
	\centering
    \subcaptionbox{  \label{fig:LzEchi+ve-Asymp} }{
\includegraphics[width=\columnwidth]{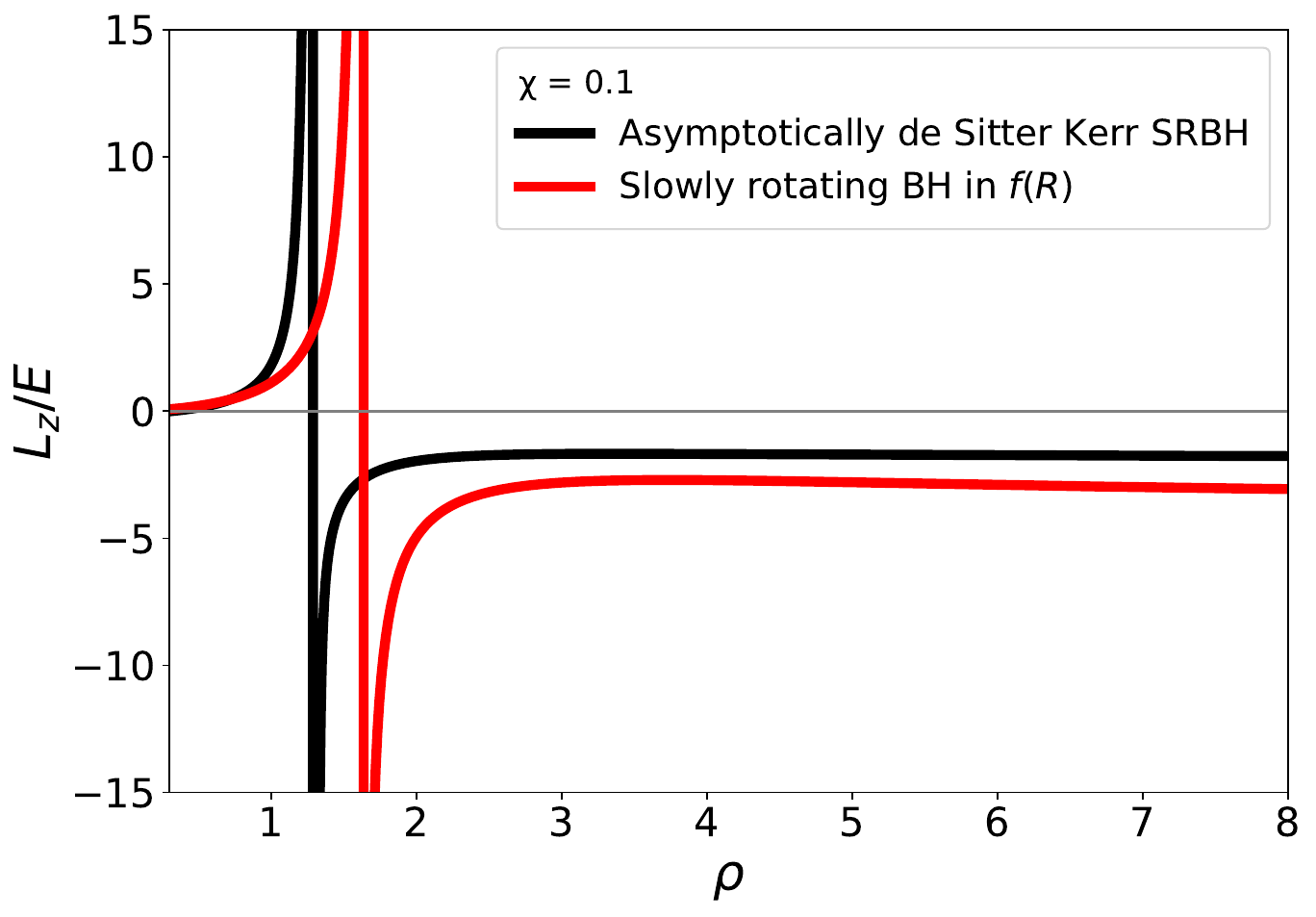}   }

\end{minipage}\hfill
\begin{minipage}[b]{.5\textwidth}
	\centering
    \subcaptionbox{ \label{fig:LzEchi-ve-Asymp} }{
\includegraphics[width=\columnwidth]{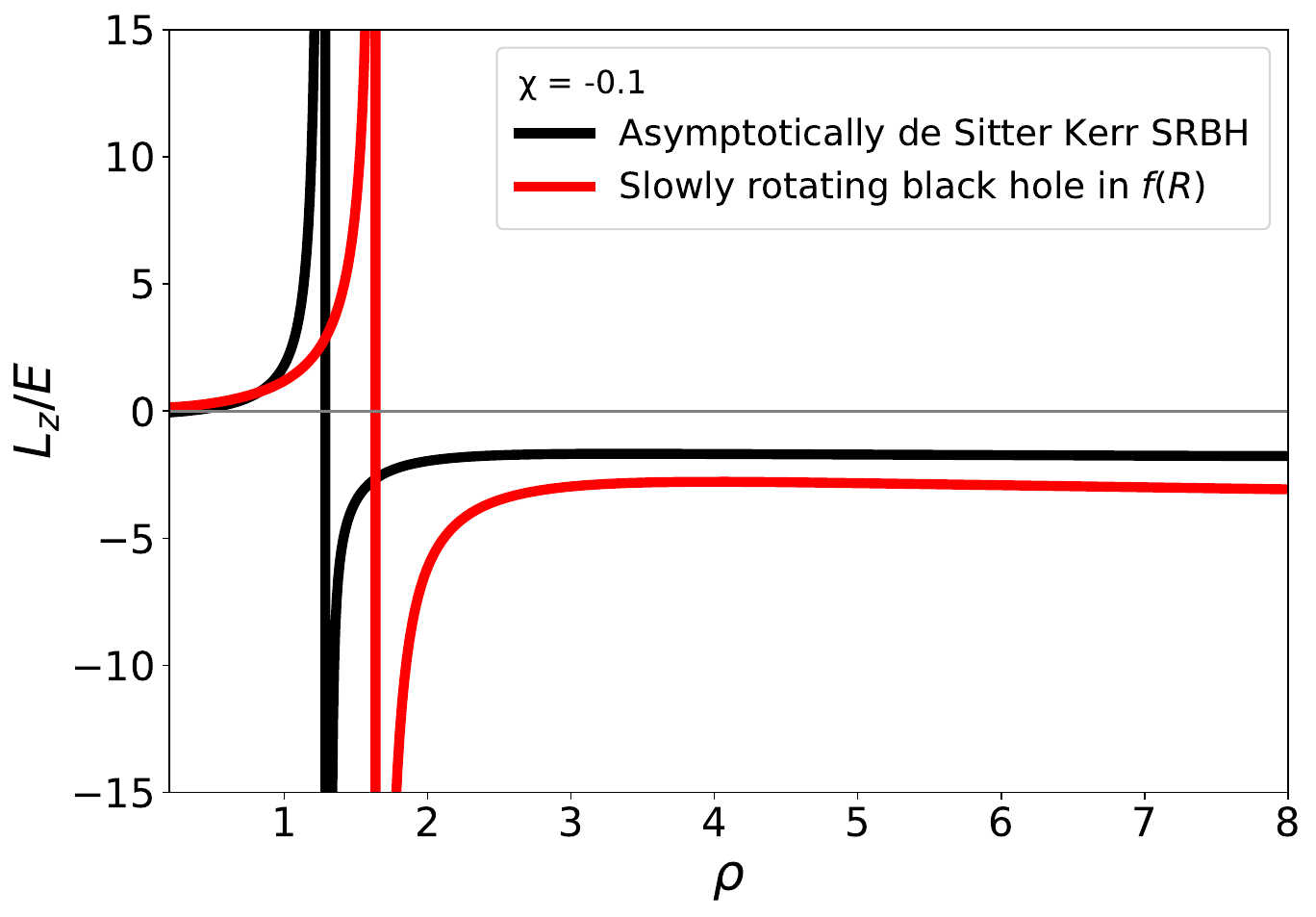}  }

\end{minipage} 
\caption{Plot of $L_z/E$ of a test particle in equatorial circular orbits as a function of $\rho$ for $\kappa^2 =1$ and $C_1 =2$. This is compared with asymptotic de Sitter Kerr SRBH \eqref{Asymp_Slow_Kerr} where $\Lambda =1$. (a) For  $\chi = 0.1$. (b) For  $\chi = - 0.1$.}
\label{fig:LzE-Asymp}
\end{figure*}

%
%
%
%

\begin{figure*}[!htb]
\begin{minipage}[b]{.5\textwidth}
	\centering
    \subcaptionbox{\label{fig:EffV-Asymp} }{
\includegraphics[width=\columnwidth]{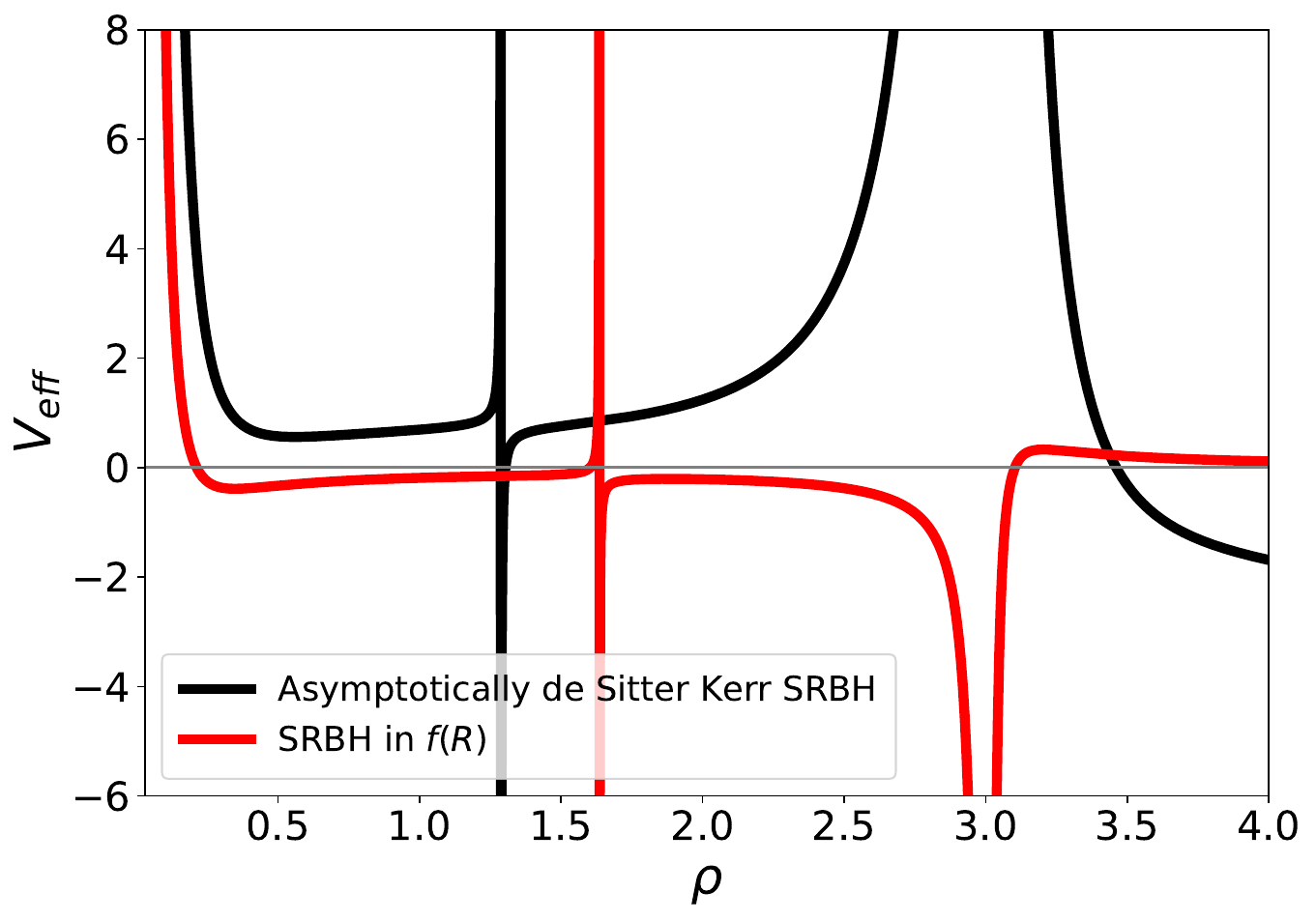}   
 }
\end{minipage}\hfill
\begin{minipage}[b]{.5\textwidth}
	\centering
    \subcaptionbox{ \label{fig:zamo-Asymp} }{
\includegraphics[width=\columnwidth]{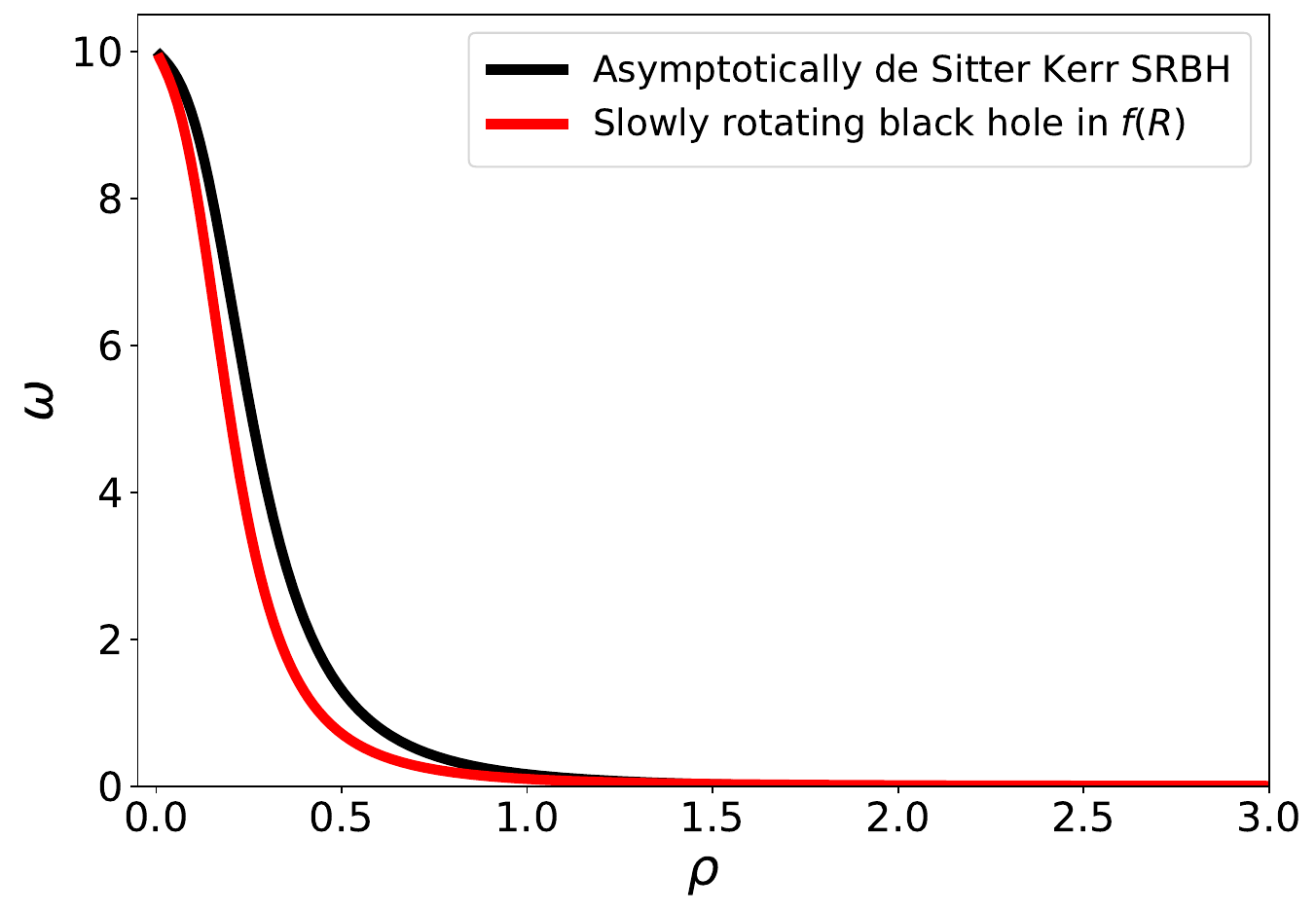}   
	}
\end{minipage} 
\caption{(a) Plot of $V_{\rm eff}$ of a test particle in the equatorial plane as a function of $\rho$. We set $C_1 = 2$ and $\Lambda =1$.
(b) Plot of the equatorial plane ZAMO's angular velocity as a function of $\rho$. For line-element  \eqref{Slow_Kerr}, we have set $\kappa^2 = 1, \chi = 0.1$ and $C_3 =1$. This is compared with asymptotic de Sitter Kerr SRBH \eqref{Asymp_Slow_Kerr} where $\Lambda =1$.}
\end{figure*}

\vspace*{12pt}

The plot of the ergosphere of the asymptotic de Sitter Kerr Eq.~(\ref{Asymp_Slow_Kerr}) is given in Fig. \eqref{fig:ergosphereNonflat}, which has similarities to Fig.~\eqref{sym}. Let us now look at the kinematic properties of both the solutions. We have compared $L_z/E$ for both positive and negative values of $\chi$ in Fig.~\eqref{fig:LzEchi+ve-Asymp} and Fig.~\eqref{fig:LzEchi-ve-Asymp} respectively. Similarly, we also shown $V_{eff}$ Fig.~\eqref{fig:EffV-Asymp} and ZAMO's angular velocity Fig~\eqref{fig:zamo-Asymp} for both the solutions. We can see that the solution Eq.~(\ref{Asymp_Slow_Kerr}) and our solution Eq.~(\ref{Slow_Kerr}) have  very similar properties.

\newcommand{\newblock}{}
\raggedright{}
\nocite{apsrev41Control}
\bibliographystyle{apsrev4-1}
%

\end{document}